\documentclass[aip,jcp,reprint,onecolumn]{revtex4-2}
\usepackage[english]{babel}
\usepackage{amsmath,amsthm,latexsym,amssymb,amsfonts,epsfig}
\usepackage{bm}
\usepackage{dsfont}
\usepackage[left = 1.35cm, right = 1.35cm, top = 2cm, bottom = 2cm]{geometry}

\usepackage{graphicx}
\usepackage{color}

\definecolor{CarmineRed}{rgb}{1.0, 0.0, 0.22}
\definecolor{OliveGreen}{rgb}{0.1, 0.4, 0.1}
\definecolor{AppleGreen}{rgb}{0.55, 0.71, 0.0}

\usepackage{setspace}

\emergencystretch=\maxdimen
\allowdisplaybreaks

\usepackage{mathtools}
\usepackage{mathrsfs}

\usepackage{type1ec} %

\usepackage{amsbsy}

\usepackage[colorlinks=true,citecolor=OliveGreen,linkcolor=OliveGreen,urlcolor=OliveGreen]{hyperref}

\newcommand{\bnabla}{\boldsymbol{\nabla}}

\newcommand{\ch}{\operatorname{ch}}
\newcommand{\sh}{\operatorname{sh}}

\newcommand{\csh}{\operatorname{csh}} 
\newcommand{\sch}{\operatorname{sch}} 

\newcommand{\sgn}{\operatorname{sgn}}

\begin{document}

\title{Hydrodynamic flows induced by localized torques (rotlets) in wedge-shaped geometries}

\author{Abdallah Daddi-Moussa-Ider}
\email{abdallah.daddi-moussa-ider@open.ac.uk}
\thanks{corresponding author.}
\affiliation{School of Mathematics and Statistics, The Open University, Walton Hall, Milton Keynes MK7 6AA, United Kingdom}

\author{Jakob Mihatsch}
\affiliation{Institut f\"ur Physik, Otto-von-Guericke-Universit\"at Magdeburg, Universit\"atsplatz 2, 39106 Magdeburg, Germany}

\author{Michael J.\ Mitchell}
\affiliation{School of Mathematics and Statistics, The Open University, Walton Hall, Milton Keynes MK7 6AA, United Kingdom}

\author{Elsen Tjhung}
\affiliation{School of Mathematics and Statistics, The Open University, Walton Hall, Milton Keynes MK7 6AA, United Kingdom}

\author{Andreas M.\ Menzel}
\email{a.menzel@ovgu.de}

\affiliation{Institut f\"ur Physik, Otto-von-Guericke-Universit\"at Magdeburg, Universit\"atsplatz 2, 39106 Magdeburg, Germany}

\begin{abstract}
Wedge-shaped geometries in low-Reynolds-number flows are of increasing importance, for instance, in the design of microfluidic devices. The corresponding Green’s functions describing the induced flow in response to a locally applied force were derived some time ago. To achieve a complete characterization of particle motion at low Reynolds numbers, we  derive the flow response to locally applied torques. This is accomplished through a direct calculation based on the Fourier–Kontorovich–Lebedev transform using the Papkovich–Neuber representation of the hydrodynamic fields. We then illustrate the resulting flow fields, highlighting their structure, key features, and dependence on the geometry and orientation of the applied torque. Based on these solutions, we compute the corresponding hydrodynamic mobility tensor that couples torque and motion. Owing to the broken spatial symmetry imposed by the wedge-shaped confinement, a particle subjected to a torque will experience not only rotational motion but also translational motion. These results provide analytical tools relevant for predicting and controlling particle behavior in confined microfluidic environments.

\end{abstract}

\maketitle

\section{Introduction}

Wedge-shaped geometries have been studied in hydrodynamics because of their peculiar effect on intrinsic fluid flows, and because of their practical abundance. 
If a net velocity is prescribed on one surface of the wedge in radial direction under low Reynolds number conditions, vortices will be induced near the edge of the wedge, specifically at sufficiently low opening angles \cite{moffatt1964viscous}. For instance, such net motion along one surface can be introduced by radial motion of the surface under no-slip conditions. Wedge-shaped geometries were further investigated under low-Reynolds-number conditions concerning their induced flows due to chemically active surfaces. It was demonstrated that resulting chemical concentration gradients can induce corresponding vortices near the edge of the wedge \cite{nowak2025diffusiophoretic}. 
More recently, the self-diffusiophoretic motion of a catalytically active spherical particle confined within a wedge-shaped domain has been examined~\cite{daddi2026toward}.

Generally, any microfluidic channel bounded by straight walls features wedge-like shapes near the edges of confinement, even if simply of the specific angle of ninety degrees. Additionally, wedges of different opening angles were introduced in microfluidic devices, for example, to promote cell sorting, also with medical applications in mind \cite{yang2018wedge}. 
Moreover, wedge-shaped microchannels were produced to generate sharp-edged microparticles \cite{zhou2022fabrication}. 

In view of the still growing field of microfluidics \cite{battat2022outlook,nunes2022introduction}, for example, if thinking of lab-on-a-chip devices, an explicit understanding and quantitative description of underlying flows and mechanisms of such geometries is thus highly desirable. If the flow is driven by internal force application, then the corresponding Green's function provides the fundamental solution from which all cases of flow can be formulated. Specifically, it quantifies the induced flow in response to a force applied locally at one point within the fluid. 

Low-Reynolds-number flows of incompressible viscous fluids near three-dimensional corners were first investigated by Sano and Hasimoto~\cite{sano76, sano1977slow, sano1978effect, hasimoto80, sano1977slow_thesis}. The corresponding Green’s function for free-slip wedge-shaped confinement has also been derived for wedges with commensurate opening angles, employing an image method~\cite{sprenger2023microswimming}.
Stokes flow near a corner has been revisited using a complex analysis approach~\cite{dauparas2018leading, dauparas2018stokes}.
More recently, we have derived corresponding mathematical expressions in previous investigations for different boundary conditions, free-slip and no-slip \cite{daddi2025proc,daddi2025jelasticity}. 

Generally, the response to applied higher-order localized force distributions, such as force dipoles or quadrupoles, can be derived from this Green’s function, which is particularly relevant for modeling microswimmers and other active particles~\cite{spagnolie12,mathijssen2016hotspots,daddi2019frequency,sprenger2020towards,zottl2023modeling}. Application of a torque at one point within the fluid is another mode of action on the system. Here, we derive the resulting fluid flow of such a ``rotlet'' applied to the fluid explicitly. In practice, example situations of imposed rotations at small length scales comprise induced mixing by rotational components \cite{lee2009effective,owen2016rapid} or micro-rheological investigations by rotating local probes \cite{wilhelm2003rotational,schmiedeberg2005one,squires2010fluid,richter2021rotating, daddi2018slow, daddi18jcp}. Typically, such rotations are induced using magnetic probe particles in reorienting external magnetic fields \cite{huang2016buckling,kreissl2021frequency}. In our context of the Green's function, we would consider them as point-like. 

Point torques (rotlets) have broad applications in microhydrodynamics. They provide a simple model for flows generated by beating cilia near surfaces, enabling predictions of fluid transport and mixing in ciliary carpets~\cite{selvan2023point, selvan2025mass}. In active systems such as colloidal microrollers, rotlets capture torque-driven flows near walls. They allow the analysis of particle transport, enhanced mixing, and collective interactions~\cite{driscoll2017unstable, delmotte2017hydrodynamic, delmotte2019hydrodynamically}. Similarly, bacterial hydrodynamics can be modeled using rotlets to describe the flows induced by rotating flagella or cell bodies. This strategy sheds light on surface accumulation, circular trajectories, and microscale mixing in bacterial suspensions~\cite{dauparas16, lopez2014dynamics, chamolly2020direct}.

In addition to deriving the flow field generated by a point torque in a wedge, we use this solution to determine, to leading order, the associated hydrodynamic coupling and rotational mobility tensors of a small particle subjected to a torque. In an unbounded fluid, a torque acting on a spherical particle generates only rotational motion, with no coupling to translation. In contrast, the wedge geometry breaks this symmetry, so a torque applied to a particle can induce a small but finite translational motion. This translation–rotation coupling arises solely from hydrodynamic interactions with the confining walls.

We proceed as follows. In Sec.~\ref{sec:math}, we overview the mathematical formulation of the problem and introduce the Fourier-Kontorovich-Lebedev transformation, a reliable mathematical tool convenient for solving the low-Reynolds-number flow equations in a wedge-shaped confinement. We use it to solve the flow equations in Sec.~\ref{sec:greens_function_FKL}. Specifically, we distinguish between two different orientations of the rotlet, namely, applying the torque along or perpendicular to the edge of the wedge. 
Resulting flows are illustrated.
In Sec.~\ref{sec:planar} we show that taking the limit of an opening angle of $\pi$ recovers previously derived solutions for a rotlet near a planar wall. Leading-order expressions for the hydrodynamic mobilities for rotational--translational coupling and purely rotational effects due to the confining boundaries are presented in Sec.~\ref{sec:mobilities}.
We conclude in Sec.~\ref{sec:concl}.

\section{Mathematical formulation}
\label{sec:math}

\begin{figure}
    \centering
    \includegraphics[width=0.5\linewidth]{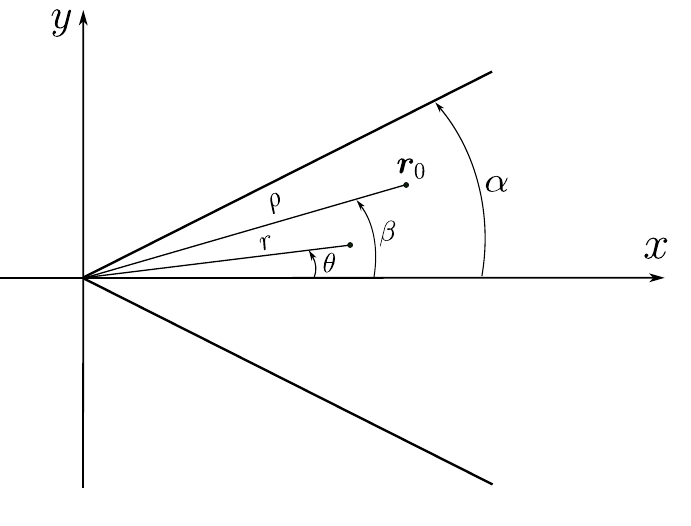}
    \caption{Schematic illustration of the system setup. A point torque singularity is located at $\bm{r}_0 = (\rho, \beta, 0)$ in the cylindrical coordinate system $(r, \theta, z)$, acting within a viscous fluid confined by two planar boundaries that form a wedge with a straight edge aligned along the $z$-axis. The bounding surfaces are situated at $\theta = \pm \alpha$.
}
    \label{fig:system-setup}
\end{figure}

We examine an incompressible, viscous fluid confined within a wedge-shaped domain whose tip is formed by a straight edge. 
The geometry of the system is depicted in Fig.~\ref{fig:system-setup}. 
A cylindrical coordinate system $(r, \theta, z)$ is employed, with the straight edge of the wedge oriented along the $z$-axis. The confined fluid is bounded by the enclosing surfaces of the wedge at $\theta = \pm\alpha$, where $\alpha \in (0, \pi/2]$. For $\alpha = \pi/2$ we obtain a semi-infinite fluid with a flat surface. The opening angle of the wedge is $2\alpha$. We consider the presence of a torque singularity (point torque or couplet) located at $(r, \theta, z) = (\rho, \beta, 0)$ in this cylindrical system, oriented in an arbitrary direction. 
We impose no-slip boundary conditions on the surfaces of the wedge.

\subsection{Governing equations} 

At low Reynolds numbers, the fluid flow is governed by the Stokes equations~\cite{kim05}
\begin{equation}
    -\boldsymbol{\nabla}p + \eta \boldsymbol{\nabla}^2 \bm{v} + 
    \tfrac{1}{2} \, \bm{L} \times  \boldsymbol{\nabla} \, \delta(\bm{r}-\bm{r_0})
    = \bm{0} \, , \quad
    \boldsymbol{\nabla} \cdot \bm{v} = 0 \, ,
    \label{eq:Stokes_Eqns}
\end{equation}
where $\bm{v}$ and $p$ represent the hydrodynamic velocity and pressure fields, respectively,
$\eta$ is the shear viscosity, 
and $\bm{L}$ is the strength of the rotlet singularity located at $\bm{r}_0$.
Our scope is to calculate the flow induced by this rotlet flow.

In an unbounded fluid, the flow field generated by a rotlet singularity is expressed as
\begin{equation}
    \bm{v}^\infty  = \frac{1}{8\pi\eta} 
    \frac{\bm{L} \times \bm{s}}{s^3} \, , 
    \label{eq:unbounded}
\end{equation}
where $\bm{s} = \bm{r}-\bm{r}_0$ and $s = |\bm{s}|$ denotes the distance between the evaluation point and the position of the torque singularity. We express it in cylindrical coordinates as
\begin{equation}
    s = \sqrt{r^2+\rho^2-2\rho r \cos(\theta-\beta)+z^2} \, . \label{eq:s}
\end{equation}
For later reference, we define the abbreviation
\begin{equation}
\bm{q} = \frac{\bm{L}}{16\pi\eta} \, , \label{eq:q-torque}
\end{equation}
which has the dimension of (length)$^{3}$(time)$^{-1}$.

A general solution to Eq.~\eqref{eq:Stokes_Eqns} is given in the Papkovich–Neuber representation as~\cite{Papkovich1932,*Papkovich1932french,Neuber1934, tran1982general}
\begin{equation}
    \bm{v} (\bm{r}) = \bnabla \bigl( \bm{r} \cdot \boldsymbol{\Phi} (\bm{r}) + \rho \, \Phi_w (\bm{r}) \bigr) - 2 \, \boldsymbol{\Phi} (\bm{r}) \, , 
    \label{eq:Papkovich}
\end{equation}
where $\boldsymbol{\Phi} (\bm{r})$ has components $\Phi_x$, $\Phi_y$, and $\Phi_z$ in the Cartesian coordinate system.
From this point onward, we omit explicitly indicating the dependence of $\bm{v}$ and $\boldsymbol{\Phi}$ on~$\bm{r}$.
Here, $\Phi_j$ are harmonic functions satisfying Laplace’s equation, $\Delta \Phi_j = 0$, for $j \in \{x, y, z, w\}$.
We note that the components of $\boldsymbol{\Phi}$ in polar coordinates are related to those in Cartesian coordinates through
$\Phi_r = \Phi_x\cos\theta + \Phi_y\sin\theta$ and
$\Phi_\theta = -\Phi_x \sin\theta + \Phi_y\cos\theta$.
Using the cylindrical coordinate system we defined, we express the velocity field in terms of its radial, azimuthal, and axial components, denoted by $v_r$, $v_\theta$, and $v_z$, respectively.
Applying the standard Cartesian-to-cylindrical coordinate transformations to Eq.~\eqref{eq:Papkovich}, the velocity components in the cylindrical system are given by
\begin{equation}
   v_r = \frac{\partial \Pi}{\partial r} - 2\Phi_r \, , \quad
   v_\theta = \frac{1}{r} \frac{\partial \Pi}{\partial \theta} - 2\Phi_\theta \, , \quad
   v_z = \frac{\partial \Pi}{\partial z} - 2\Phi_z \, , 
   \label{eq:velocity_field}
\end{equation}
where
\begin{equation}
    \Pi = r\Phi_r + z\Phi_z + \rho \Phi_w \, .
    \label{eq:Pi}
\end{equation}

In our wedge-shaped geometry, the general solution of the flow equations can be expressed as
\begin{equation}
    \Phi_j = \phi_j^\infty + \phi_j \, , 
\end{equation}
for $j\in \{x,y,z,w\}$, with $\phi_j^\infty$ representing the harmonic function in an infinitely extended fluid medium, which we refer to hereinafter as the ``free-space" solution derived from Eq.~\eqref{eq:unbounded}.
The auxiliary fields $\phi_j$ are unknown harmonic functions, also referred to as the complementary solution.
By applying no-slip boundary conditions at the surfaces of the wedge, Eqs.~\eqref{eq:velocity_field} imply that a natural choice 
at $\theta = \pm\alpha$ is
\begin{equation}
    \Phi_r = 0 \, , \quad
    \Phi_z = 0 \, , \quad
    \Phi_w = 0 \, , \quad
    \frac{1}{r}
    \frac{\partial \Pi}{\partial\theta}  - 2\Phi_\theta = 0 \,  \qquad \text{at}~ \theta = \pm\alpha \, .
    \label{eq:BCs_real}
\end{equation}
Accordingly, we set $\Pi = 0$ at $\theta = \pm\alpha$, so that $\partial\Pi/\partial r = \partial\Pi/\partial z = 0$ at the boundary, while $\partial\Pi/\partial\theta$ is not necessarily zero.

\subsection{Fourier-Kontorovich-Lebedev transform}

\subsubsection{General overview and definitions}

To determine the velocity field generated by a point torque in a wedge-shaped geometry, we employ the Fourier–Kontorovich–Lebedev (FKL) transform. In this approach, the axial $z$, and radial $r$ coordinates, are transformed into their corresponding wavenumbers, denoted by $k$ and $\nu$, respectively.
The transform was first introduced by the Soviet mathematicians Kontorovich and Lebedev to address some boundary value problems~\cite{kontorovich1938one, kontorovich1939method,kontorovich1939application}. The mathematical framework was later developed by Lebedev~\cite{lebedev1946, lebedev1949}. We refer the reader to Erd{\'e}lyi~\textit{et~al.}~\cite[p.~75]{erdelyi1953higher} for additional details on this integral transform.

Laplace's equation becomes significantly simpler to solve in FKL space because the transform diagonalizes the differential operators involved, reducing the problem to an algebraic form in the transformed variables. The Fourier transform applied along the axial direction converts the associated derivative into an algebraic multiplier. At the same time, the Kontorovich–Lebedev transform, which is naturally suited to Bessel-type operators, diagonalizes the planar part of the Laplacian. As a result, the full partial differential equation, which in physical space involves coupled derivatives and radial dependence, reduces in FKL space to a straightforward ordinary differential equation in the polar angle. The solution can then be obtained directly in FKL space and recovered in physical space by applying the inverse transforms.

We begin by defining the forward Fourier transform of a function $f(r,z)$ with respect to $z$ as
\begin{equation}
    \hat{f}(r, k) 
    := \mathscr{F} \left\{ f \right\}
    = 
    \int_{-\infty}^\infty f(r,z) \, e^{ikz} \, \mathrm{d} z \, , 
    \label{eq:Fourier}
\end{equation}
where we adopt the convention of a plus sign in the exponent. We then define the Kontorovich–Lebedev (KL) transform with respect to $r$ of the already Fourier-transformed quantity as
\begin{equation}
    \widetilde{f}(\nu, k) 
    := \mathscr{K}_{i\nu} \left\{ \hat{f} \right\}
    = \int_0^\infty \hat{f}(r, k) \, K_{i\nu} (|k|r) \, r^{-1} \, \mathrm{d} r  \, .
    \label{eq:KL}
\end{equation}
Here, $K_{i\nu}(|k|r)$ is the modified Bessel function of the second kind~\cite{abramowitz72} with purely imaginary order $i\nu$. 
For positive arguments, $K_{i\nu}$ is real if $\nu$ is real.
Throughout, the symbol of a hat~~$\widehat{}$~~denotes the Fourier transform of the original function, while a tilde~~$\widetilde{}\,$~~indicates the Kontorovich–Lebedev transform of the Fourier-transformed function. We note that the polar angle~$\theta$ remains unchanged under these transformations. As a result, the partial differential equation in Eq.~\eqref{eq:Stokes_Eqns}, which governs the hydrodynamic velocity field, is reduced to a system of ordinary differential equations in the polar angle~$\theta$.

In what follows, the hyperbolic sine, cosine, and tangent functions are denoted by sh, ch, and th, respectively.
In addition, sch and csh are the secant and cosecant hyperbolic functions, respectively.
For convenience, we define the notation for the combined FKL transform as
\begin{equation}
    \widetilde{f} = \mathscr{T}_{i\nu} \left\{ f \right\}
    = \mathscr{K}_{i\nu} \left\{ \mathscr{F} \left\{ f \right\} \right\} .
\end{equation}

Generally, the FKL transform can be written as a double infinite integral over $r$ and $z$. Typically, there is no prescribed order of integration. In practice, however, it is common to first perform the forward Fourier transform, followed by the forward Kontorovich–Lebedev transform.

The inverse FKL transform is expressed as the double integral
\begin{align}
    f(r,z) &=  
    \mathscr{T}_{i\nu}^{-1} \left\{ \widetilde{f} \right\}
    = \frac{1}{\pi^3} 
    \int_{-\infty}^\infty \mathrm{d}k \, e^{-ikz} 
    \int_0^\infty \widetilde{f}(\nu, k) \, K_{i\nu} (|k|r) \sh (\pi\nu) \, \nu \, \mathrm{d} \nu \, . \label{eq:inverse_FKL_transform}
\end{align} 

It is worth noting that the forward FKL transform is typically performed by first computing the Fourier transform, as defined in Eq.~\eqref{eq:Fourier}, followed by the Kontorovich–Lebedev (KL) transform given in Eq.~\eqref{eq:KL}. In contrast, the inverse transformation is generally not carried out by applying the inverse KL transform first; instead, the inverse Fourier transform is evaluated initially. This approach reduces the problem to a single infinite integral over the radial wavenumber~$\nu$, which is usually computed numerically.

\subsubsection{Key Properties of the FKL Transform}
\label{subsubsec:prop}

Knowing the FKL transform of $1/s$, the corresponding FKL transforms of the infinite-space hydrodynamic solution in Eq.~\eqref{eq:unbounded} can be derived by taking appropriate derivatives and applying the relevant properties of FKL transforms. While Fourier transforms are generally straightforward to evaluate, determining KL transforms often requires consulting classical references of tabulated integrals and transforms, such as Erdélyi \textit{et al.}~\cite{erdelyi1953higher}.

The FKL transform of the derivative with respect to $z$ can be directly obtained using the FKL transform properties, analogous to the derivative property of the Fourier transform,
\begin{equation}
    \mathscr{T}_{i\nu} \left\{ \frac{\partial f}{\partial z} \right\} = -ik \mathscr{T}_{i\nu} \{f\} \, .
    \label{eq:FKL-property-diff-Z}
\end{equation}

The following identities are proven in the Appendix.
The FKL transform of the derivative with respect to $r$ can be readily obtained using the property
\begin{equation}
    \mathscr{T}_{i\nu} \left\{ \frac{\partial f}{\partial r} \right\}
    = \frac{|k|}{2i\nu} 
    \left( (i\nu+1) \mathscr{T}_{i\nu+1} \{f\}
    +  (i\nu-1) \mathscr{T}_{i\nu-1} \{f\}
    \right) .
    \label{eq:FKL-property-diff-R}
\end{equation}
Moreover, the FKL transform of a function divided by $r$ is given by
\begin{equation}
    \mathscr{T}_{i\nu} \left\{ \frac{f}{r} \right\}
    = \frac{|k|}{2i\nu} \left( \mathscr{T}_{i\nu+1} \{f\} - \mathscr{T}_{i\nu-1} \{f\} \right) .
    \label{eq:FKL-property-division-by-r}
\end{equation}
Furthermore, the following FKL transform property holds,
\begin{equation}
    \mathscr{T}_{i\nu} \left\{ \frac{z}{r} \, f \right\} 
    = {} -\frac{\sgn k}{2\nu} 
    \left(
    \left(
    i\nu + 1 + k\, \frac{\partial}{\partial k} \right) \mathscr{T}_{i\nu+ 1} \left\{ f\right\}
    + \left(
    i\nu - 1 - k\, \frac{\partial}{\partial k} \right) \mathscr{T}_{i\nu-1} \left\{ f\right\}
    \right) ,
    \label{eq:FKL_rzf}
\end{equation}
where sgn denotes the sign function.

It is worth noting that when $f$ is an even function of $z$, as is the case with $1/s$, its FKL transform is real. As a result, for transforms that preserve $z$-parity, the FKL transform remains real, and the two terms in Eqs.~\eqref{eq:FKL-property-diff-R} and \eqref{eq:FKL-property-division-by-r} are complex conjugates, so the transform can be obtained by taking twice the real part of either term. In contrast, for transforms that reverse $z$-parity, the transform of an even function yields a purely imaginary result. Consequently, the two terms in Eqs.~\eqref{eq:FKL-property-diff-Z} and \eqref{eq:FKL_rzf} are negative complex conjugates, with identical imaginary parts and real parts equal in magnitude but opposite in sign.
However, for the sake of streamlined subsequent calculations, we choose to present the results in the above forms.

Besides, many previous studies employed Mellin transforms to solve boundary-value problems in wedge-shaped domains~\cite{tranter1948, martin2017}. The Mellin transform is particularly effective when describing flows in two-dimensional corners, as outlined by Moffatt~\cite{moffatt1964viscous, moffatt1964}. 
For example, recently, this transform has been used to analyze flows induced in a two-dimensional corner by the chemical activity of the confining boundaries~\cite{nowak2025diffusiophoretic}. 
Both the Mellin transform and the here-employed Kontorovich–Lebedev transform can be applied to the radial coordinate in the governing equations. However, the Mellin transform is often more suitable for two-dimensional problems, while the KL transform is often better suited for three-dimensional problems. These transforms reduce the Laplace equation to an ordinary differential equation in the polar angle. In the three-dimensional case, the Fourier–Kontorovich–Lebedev transform is obtained by additionally applying a Fourier transform.

\section{Solution in FKL space and real space}
\label{sec:greens_function_FKL}

As discussed earlier, the solution for the velocity field can be expressed as a linear superposition of the free-space solution and a complementary solution, with the latter ensuring that the boundary conditions are satisfied.
Depending on the orientation of the point torque, the free-space solution must be expressed in FKL space to determine the coefficients of the image solution. These FKL transforms can be obtained using the properties outlined in Sec.~\ref{subsubsec:prop}.

\subsection{Free-space solution}

\begin{table}[]
    \centering
    {
\renewcommand{\arraystretch}{3}
\setlength{\tabcolsep}{12pt}
    \begin{tabular}{|c|c|}
    \hline
    $f$ & $\displaystyle \mathscr{T}_{i\nu} \{f\} =  
    \int_{-\infty}^\infty \mathrm{d}z \, e^{ikz}
    \int_0^\infty f(r,z) K_{i\nu} (|k|r) \, r^{-1} \, \mathrm{d} r $ \\[3pt]
    \hline
    \hline
    $ \cfrac{\partial}{\partial z} \cfrac{1}{s}$
    & 
    $\cfrac{2\pi}{i\nu} \, \cfrac{\ch(a\nu)}{\sh(\pi\nu)} \, k \, K_{i\nu} (|k|\rho) $ \\[5pt]
    \hline
     $ \cfrac{\partial}{\partial r} \cfrac{1}{s} $ 
     & 
    $\cfrac{\pi}{\nu} \, |k|
    \left( 
     \cfrac{\ch \left( a(\nu-i)\right)}{\sh\left( \pi(\nu-i)\right)} \, K_{i\nu+1} (|k|\rho) 
     + \cfrac{\ch \left( a(\nu+i)\right)}{\sh\left( \pi(\nu+i)\right)} \, K_{i\nu-1} (|k|\rho) 
    \right) $ \\[5pt]
    \hline
    $ \cfrac{1}{r} \cfrac{\partial}{\partial a} \cfrac{1}{s} $
    &  $\cfrac{\pi}{i\nu} \, |k|
    \left( 
     \cfrac{\sh \left( a(\nu-i)\right)}{\sh\left( \pi(\nu-i)\right)} \, K_{i\nu+1} (|k|\rho) 
     - \cfrac{\sh \left( a(\nu+i)\right)}{\sh\left( \pi(\nu+i)\right)} \, K_{i\nu-1} (|k|\rho) 
    \right) $ \\[3pt]
    \hline
    \hline
    \end{tabular}
    }
    \caption{Elementary FKL transforms used in deriving the FKL transform of the free-space contribution to the solution of the hydrodynamic equations for a point torque.
    Here, $s$ represents the distance from the point torque singularity, as defined in cylindrical coordinates by Eq.~\eqref{eq:s}, and $a=\pi-|\theta-\beta|$.
    Since $s$ is an even function of $z$, the first row of transforms yields a purely imaginary value, whereas the second and third entries are real.
    }
    \label{tab:FKL-properties}
\end{table}

To compute the FKL transform of the fluid velocity in an infinite fluid medium, we first require the FKL transform of $1/s$.
Then, the free-space rotlet can be expressed in terms of the partial derivatives of $1/s$.
The forward Fourier transform can readily be obtained as 
\begin{equation}
    \mathscr{F} \left\{ \frac{1}{s} \right\} = 
    2 K_0 \left( |k|R\right) , 
\end{equation}
where we have defined
\begin{equation}
    R=s(z=0) = \sqrt{r^2+\rho^2-2\rho r \cos(\theta-\beta)} \, .
\end{equation}
We next employ the Table of Integral Transforms by Erd{\'e}lyi~\textit{et~al.}~\cite[p.~175]{erdelyi54} to derive the FKL transform of $1/s$ as
\begin{equation}
    \mathscr{T}_{i\nu} \left\{ \frac{1}{s} \right\}
    = \frac{2\pi}{\nu \sh(\pi\nu)} \ch\bigl( \left( \pi - |\theta - \beta| \right) \nu\bigl) \, K_{i\nu} (|k|\rho) \, .
    \label{eq:FKL-1-over-s}
\end{equation}
For convenience, we now introduce the abbreviation $a = \pi - |\theta - \beta|$, so that $\cos(\theta-\beta) = -\cos a$ and $\sin(\theta-\beta) = \sgn(\theta-\beta) \sin a$, with $\sgn$ the sign function.
The following identities hold, 
\begin{equation}
\frac{z}{s^3} = -\frac{\partial}{\partial z} \frac{1}{s} \, , 
\quad
    \frac{1}{s^3} \left( \rho\cos(\theta-\beta)-r \right)
    = \frac{\partial}{\partial r} \frac{1}{s} \, , 
    \quad
    \frac{\rho}{s^3} \,\sin(\theta-\beta)
    = \sgn(\theta-\beta) \, \frac{1}{r} \frac{\partial}{\partial a} \frac{1}{s} \, ,
    \quad
    \frac{r\cos(\theta-\beta)-\rho}{s^3} = \frac{\partial}{\partial \rho} \frac{1}{s} \, .
    \label{eq:Free-Space-identities}
\end{equation}
As we will see, the free-space fluid velocity can be obtained from these identities. 

Using Eqs.~\eqref{eq:FKL-property-diff-Z}–\eqref{eq:FKL-property-division-by-r} together with Eq.~\eqref{eq:FKL-1-over-s}, Tab.~\ref{tab:FKL-properties} presents the elementary transforms to determine the FKL transform of identities given by Eq.~\eqref{eq:Free-Space-identities}.
We remark that the FKL transform of the derivative of a function with respect to~$a$ is simply obtained by differentiating the FKL transformed function with respect to~$a$, because $a$ depends on the polar angle and is not affected by the transformation.

\subsection{Complementary solution}

Starting from the transformed representation of the governing operators, the Laplace equation simplifies considerably under the FKL transform.
In FKL space, the Laplace equation $\boldsymbol{\nabla}^2 f = 0$ takes the form~\cite{daddi2025proc}
\begin{equation}
    \left( \frac{\partial^2}{\partial\theta^2} - \nu^2 \right) \widetilde{f} = 0 \, ,
\end{equation}
resulting in a homogeneous second-order ordinary differential equation for the transformed function.

Since the functions $\phi_j$, $j \in \{x, y, z, w\}$, are harmonic satisfying the Laplace equation, the image solution in FKL space can be expressed as
\begin{equation}
    \widetilde{\phi}_j = A_j \sh(\theta\nu) + A_j^\dagger \ch(\theta\nu) \, , 
    \label{eq:solution_form_image}
\end{equation}
where $A_j$ and $A_j^\dagger$ are coefficients to be determined by imposing the boundary conditions at $\theta = \pm \alpha$. 

We will demonstrate that the axial torque satisfies $A_z = A_z^\dagger = 0$.
For $j \in \{x,y,w\}$ in the case of an axial torque, and for $j = z$ in the case of transverse torque, the coefficients can be expressed in the form
\begin{equation}
    A_j = \frac{2\pi q_l}{\nu\sh(\pi\nu)}
    \left( \frac{\Lambda_j}{\rho} + \mathrm{H}_j \, \frac{\partial}{\partial\rho} \right)
    K_{i\nu} (|k|\rho) \, , 
    \qquad
    A_j^\dagger = \frac{2\pi q_l}{\nu\sh(\pi\nu)}
    \left( \frac{\Lambda_j^\dagger}{\rho} + \mathrm{H}_j^\dagger \, \frac{\partial}{\partial\rho} \right)
    K_{i\nu} (|k|\rho) \, , 
    \label{eq:CoeffLamOmega}
\end{equation}
where $l\in \{\parallel, \perp\}$ denoting the axial or transverse components of the point torque given in Eq.~\eqref{eq:q-torque}.
Here, the coefficients $\Lambda_j$, $\mathrm{H}_j$, $\Lambda_j^\dagger$, and $\mathrm{H}_j^\dagger$ are functions of $\nu$, $\alpha$, and $\beta$, and, in the case of transverse torque, also depend on $\delta$.

For transverse torque, and for $j \in \{x, y, w\}$, the coefficients are expressed as
\begin{equation}
    A_j = \frac{2\pi q_\perp}{\nu\sh(\pi\nu)} \, ik \, \Delta_j K_{i\nu}(|k|\rho) \, .
    \qquad
    A_j^\dagger = \frac{2\pi q_\perp}{\nu\sh(\pi\nu)} \, ik \, \Delta_j^\dagger K_{i\nu}(|k|\rho) \, ,
    \label{eq:CoeffDelta}
\end{equation}
which is expressed in terms of $\Delta_j$ and $\Delta_j^\dagger$. They are functions of $\nu$, $\alpha$, $\beta$, and $\delta$.
By expressing the coefficients in this form, the dependence of $A_j$ and $A_j^\dagger$ on $k$ becomes explicit. This step facilitates integration with respect to $k$ during the inverse transforms.
The reason for including $\nu \sh(\pi)$ in the denominator of these expressions is to simplify the resulting formulas upon performing the inverse transformation; see Eq.~\eqref{eq:inverse_FKL_transform}.

In the following, we consider each orientation of the torque separately, starting with a rotlet directed along the axial direction parallel to the edge of the wedge (axial torque). Afterwards, we examine cases of rotlet orientation perpendicular to the edge of the wedge (transverse torque), with point torques oriented along the radial or azimuthal directions.
In each case, we begin by writing the expressions for the free-space components and then determine the complementary solution to ensure that the no-slip boundary conditions are satisfied on the surfaces of the wedge.

\subsubsection{Rotlet oriented in the axial direction}
\label{subsubsec:AXIAL}

In an infinite medium, the velocity field resulting from a point torque oriented 
along the $z$-direction 
is obtained as
\begin{equation}
    \phi_r^\infty = \frac{q_\parallel}{s^3} \, \rho \sin(\theta-\beta) \, , \quad
    \phi_\theta^\infty = \frac{q_\parallel}{s^3} \left( \rho\cos(\theta-\beta)-r \right) \, , \quad
    \phi_z^\infty = 0 \, , \quad
    \phi_w^\infty = -\frac{q_\parallel}{s^3} \, r \sin(\theta-\beta) \, . \label{eq:AXIAL_bulk}
\end{equation}
Applying the identities in Eq.~\eqref{eq:Free-Space-identities}, we arrive at
\begin{equation}
    \widetilde{\phi}_r^\infty = q_\parallel \sgn (\theta-\beta) \,
    \mathscr{T}_{i\nu} \left\{ \frac{1}{r} \frac{\partial}{\partial a} \frac{1}{s} \right\} , \quad
    \widetilde{\phi}_\theta^\infty = q_\parallel \, \mathscr{T}_{i\nu} \left\{ \frac{\partial}{\partial r} \frac{1}{s} \right\}  , \quad
    \widetilde{\phi}_z^\infty = 0 \, , \quad
    \widetilde{\phi}_w^\infty = -q_\parallel \sgn (\theta-\beta) \, \frac{1}{\rho} \frac{\partial}{\partial a} \mathscr{T}_{i\nu} \left\{  \frac{1}{s} \right\} ,
\end{equation}
where the elementary FKL transforms of the partial derivatives are provided in Tab.~\ref{tab:FKL-properties}.
Noting that $r\phi_r^\infty + \rho \phi_w^\infty = 0$, Eq.~\eqref{eq:BCs_real} can be expressed in FKL space as
\begin{equation}
    \widetilde{\phi}_r^\infty + \widetilde{\phi}_r = 0 \, , \quad
    \widetilde{\phi}_z = 0 \, , \quad
    \widetilde{\phi}_w^\infty + \widetilde{\phi}_w = 0 \, , \quad
    \frac{\partial \widetilde{\phi}_r}{\partial\theta}
    + \rho\, \frac{\partial}{\partial\theta} \, \mathscr{T}_{i\nu} \left\{ \frac{\phi_w}{r} \right\} 
    -2 \left( \widetilde{\phi}_\theta^\infty + \widetilde{\phi}_\theta \right) = 0 \, , \qquad \text{at}~ \theta = \pm\alpha \, . 
\end{equation}

Since the solution for $\widetilde{\phi}_z$ given by Eq.~\eqref{eq:solution_form_image} must vanish at both $\theta = \pm\alpha$, it follows that $\phi_z$ vanishes everywhere. For the evaluation of the FKL transform of $\phi_w/r$, we use Eq.~\eqref{eq:FKL-property-division-by-r}. This leaves us with a system of six linear equations for the unknown coefficients.
The solution for $\Lambda_w$ and $\Lambda_w^\dagger$ is straightforward and can be written as
\begin{equation}
    \Lambda_w = \nu\ch(\beta\nu) \sh \left( (\pi-\alpha)\nu\right) \csh(\alpha\nu) \, , \qquad\quad
    \Lambda_w^\dagger = \nu\sh(\beta\nu) \ch \left( (\pi-\alpha)\nu\right) \sch(\alpha\nu) \, .
\end{equation}
In addition, $\mathrm{H}_w = \mathrm{H}_w^\dagger = 0$.
The expressions for the remaining coefficients are considerably more complex, so we simplify them using abbreviations.
We define the shorthand notations  
\begin{equation}
    \lambda_{1} =  2\left( \sh^{2}(\alpha \nu) + \sin^{2}\alpha \right), \,
    \lambda_{2} = 2\left(\sh^{2}(\alpha \nu) + \cos^{2}\alpha\right), \,\,
    \delta_{1}^{\pm} = 2\left(\ch^{2}(\alpha \nu) \pm \cos(2\alpha)\right), \,
    \delta_{2}^{\pm} = 2\left(\sh^{2}(\alpha \nu) \pm \cos(2\alpha)\right), 
\end{equation}
together with
\begin{equation}
    \Gamma_\pm^{-1} = 
    \pm \bigl( \ch(2\alpha \nu)\pm\cos(2\alpha) \bigr) \bigl( \sh(2\alpha \nu)\mp\nu \sin(2\alpha) \bigr)  \, . \label{eq:GAMMA}
\end{equation}
Accordingly, we express the solution for $j\in \{x,y\}$ in the form
\begin{subequations}
    \begin{align}
    \Lambda_j &= \Gamma_\sigma \, \nu \left( a_j \nu \sin(2\alpha) + b_j  \right) \, , \qquad\quad
    \mathrm{H}_j =  \Gamma_\sigma \left( p_j \nu\sin(2\alpha) + w_j \right) \, , \\[3pt]
    \Lambda_j^\dagger &= \Gamma_{-\sigma} \, \nu \left( a_j^\dagger \nu \sin(2\alpha) + b_j^\dagger  \right) \, , \quad\quad
    \mathrm{H}_j^\dagger =  \Gamma_{-\sigma} \left( p_j^\dagger \nu\sin(2\alpha) + w_j^\dagger \right) \, , 
\end{align}
\end{subequations}
where $\sigma = -$ for $j = x$ and $\sigma = +$ for $j = y$.
The coefficients defining $\Lambda_j$ and $\Lambda_j^\dagger$ are obtained as
\begin{subequations}
    \begin{align}
    a_x &= \cos\beta \ch(\beta \nu) \bigl(\sh(\pi \nu) \sh(2\alpha \nu) - \lambda_1 \ch(\pi \nu)\bigr) + \sin\beta\sin(2\alpha)  \sh(\beta \nu) \sh(\pi \nu) \, ,  \\
    a_y &= \sin\beta \ch(\beta \nu) \bigl(\sh(\pi \nu) \sh(2\alpha \nu) - \lambda_2 \ch(\pi \nu) \bigr) + \cos\beta \sin(2\alpha)  \sh(\beta \nu)   \sh(\pi \nu) \, ,  \\
    a_x^\dagger &= \cos\beta \sh(\beta \nu)  \bigl( \lambda_2 \ch(\pi \nu) - \sh(\pi \nu) \sh(2\alpha \nu) \bigr) + \sin\beta \sin(2\alpha) \ch(\beta \nu)  \sh(\pi \nu)  \, ,  \\
    a_y^\dagger &= \sin\beta \sh(\beta \nu) \bigl( \lambda_1 \ch(\pi \nu) - \sh(\pi \nu) \sh(2\alpha \nu) \bigr) +  \cos\beta \sin(2\alpha) \ch(\beta \nu) \sh(\pi \nu) \, , 
\end{align}
\end{subequations}
and 
\begin{subequations}
    \begin{align}
    b_x &=  2 \ch(\alpha \nu) \sh(\pi \nu)\bigl(\delta_1^- \cos\beta \ch(\alpha \nu) \ch(\beta \nu)  - \sin\beta \sin(2\alpha) \sh(\alpha \nu) \sh(\beta \nu)   \bigr)  - \lambda_1 \cos\beta \sh(2\alpha \nu)  \ch(\beta \nu) \ch(\pi \nu) \, , \\
    b_y &= \lambda_2  \sin\beta \sh(2\alpha \nu) \ch(\beta \nu) \ch(\pi \nu)-2\ch(\alpha \nu) \sh(\pi \nu) \bigl(\delta_1^+ \sin\beta  \ch(\alpha \nu) \ch(\beta \nu) - \cos\beta  \sin(2\alpha) \sh(\alpha \nu) \sh(\beta \nu)  \bigr) \, , \\
    b_x^\dagger &= 2 \sh(\alpha \nu) \sh(\pi \nu) \bigl(\delta_2^+ \cos\beta  \sh(\alpha \nu) \sh(\beta \nu) + \sin\beta \sin(2\alpha)\ch(\alpha \nu) \ch(\beta \nu)  \bigr)  - \lambda_2 \cos\beta \sh(2\alpha \nu)  \sh(\beta \nu) \ch(\pi \nu) \, , \\
    b_y^\dagger &= \lambda_1 \sin\beta \sh(2\alpha \nu)  \sh(\beta \nu) \ch(\pi \nu)  -2\sh(\alpha \nu) \sh(\pi \nu)\bigl(\delta_2^- \sin\beta  \sh(\alpha \nu) \sh(\beta \nu) + \cos\beta \sin(2\alpha)  \ch(\alpha \nu) \ch(\beta \nu) \bigr) \, . 
\end{align}
\end{subequations}
The parameters defining $\mathrm{H}_j$ and $\mathrm{H}_j^\dagger$ are obtained as
\begin{subequations}
    \begin{align}
    p_x &= \sin\beta \sh(\beta \nu) \bigl (\sh(\pi \nu) \sh(2\alpha \nu) - \lambda_1 \ch(\pi \nu) \bigr) -  \cos\beta \sin(2\alpha) \ch(\beta \nu) \sh(\pi \nu) \, ,  \\
    p_y &= \cos\beta \sh(\beta \nu) \bigl(\lambda_2 \ch(\pi \nu) - \sh(\pi \nu) \sh(2\alpha \nu)\bigr) + \sin\beta \sin(2\alpha)\ch(\beta \nu) \sh(\pi \nu)   \, ,  \\
     p_x^\dagger &= \sin\beta \ch(\beta \nu) \bigl(\lambda_2 \ch(\pi \nu) - \sh(\pi \nu) \sh(2\alpha \nu) \bigr) - \cos\beta \sin(2\alpha) \sh(\beta \nu)  \sh(\pi \nu) \, ,  \\
    p_y^\dagger &= \cos\beta \ch(\beta \nu) \bigl(\sh(\pi \nu) \sh(2\alpha \nu) - \lambda_1 \ch(\pi \nu) \bigr) + \sin\beta \sin(2\alpha)  \sh(\beta \nu) \sh(\pi \nu)\, , 
\end{align}
\end{subequations}
and 
\begin{subequations}
    \begin{align}
    w_x &= 2 \ch(\alpha \nu) \sh(\pi \nu) \bigl(\delta_1^- \sin\beta \ch(\alpha \nu) \sh(\beta \nu)  
    + \cos\beta \sin(2\alpha) \sh(\alpha \nu) \ch(\beta \nu) \bigr) 
    - \lambda_1 \sin\beta \sh(2\alpha \nu)  \sh(\beta \nu) \ch(\pi \nu) \, , \\
    w_y &=  2 \ch(\alpha \nu) \sh(\pi \nu) \bigl(\delta_1^+ \cos\beta  \ch(\alpha \nu) \sh(\beta \nu) + \sin\beta \sin(2\alpha) \sh(\alpha \nu)  \ch(\beta \nu)  \bigr) - 
    \lambda_2  \cos\beta \sh(2\alpha \nu) \sh(\beta \nu) \ch(\pi \nu)\, ,  \\
    w_x^\dagger &=  2 \sh(\alpha \nu) \sh(\pi \nu) \bigl(\delta_2^+ \sin\beta  \sh(\alpha \nu) \ch(\beta \nu) - \cos\beta \sin(2\alpha) \ch(\alpha \nu) \sh(\beta \nu)   \bigr) 
    - \lambda_2 \sin\beta \sh(2\alpha \nu)  \ch(\beta \nu) \ch(\pi \nu) \, ,  \\
    w_y^\dagger &= 2 \sh(\alpha \nu) \sh(\pi \nu) \bigl(\delta_2^- \cos\beta \sh(\alpha \nu)  \ch(\beta \nu) - \sin\beta   \sin(2\alpha) \ch(\alpha \nu)\sh(\beta \nu)\bigr) 
    - \lambda_1 \cos\beta \sh(2\alpha \nu)  \ch(\beta \nu) \ch(\pi \nu) \, .
\end{align}
\end{subequations}

\begin{figure}
    \centering
    \includegraphics[width=0.7\linewidth]{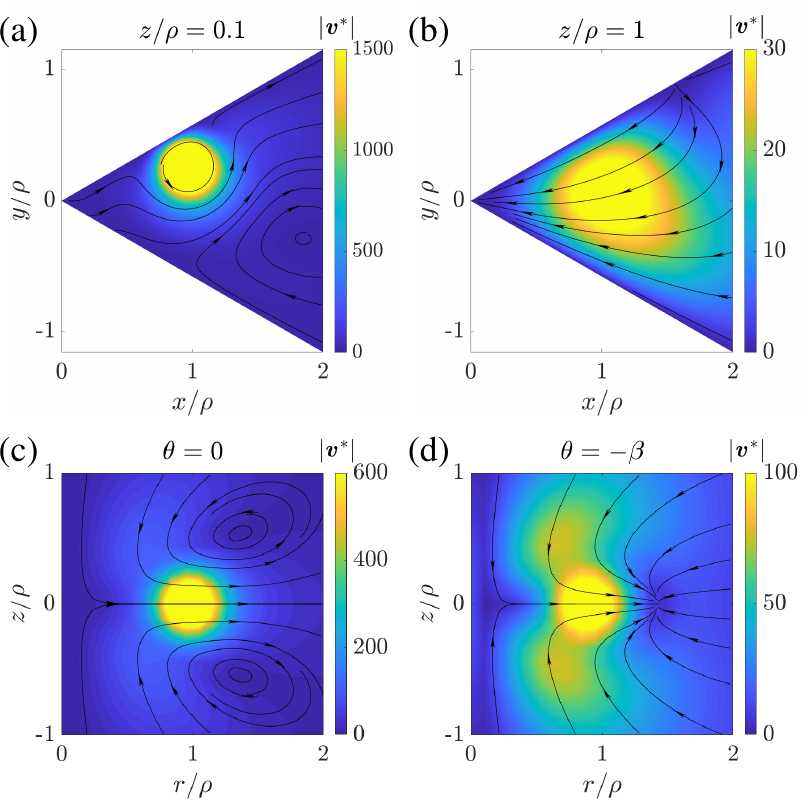}
    \caption{Streamlines and quiver plots of the viscous flow field induced by an axially oriented rotlet singularity in a wedge-like confinement with no-slip boundary conditions and semi–opening angle $\alpha = \pi/6$. The rotlet is located at $\beta = \alpha/2$, see Fig.~\ref{fig:system-setup}, at $z=0$ and points along the direction of the edge of the wedge (towards the reader). Results are presented in the radial–azimuthal plane, that is, in the plane normal to the edge of the wedge in (a) at height $z/\rho = 0.1$ and (b) height $z/\rho = 1$ above the plane of torque application. Conversely, the bottom row depicts the radial–axial plane containing the edge of the wedge for (c) $\theta = 0$ and (d) $\theta = -\beta$. The depicted scaled velocity field is defined as $\bm{v}^* = \left( \rho^2/ q_\parallel\right) \bm{v} $,
    while the plots are generated using the results derived in Sec.~\ref{subsubsec:AXIAL}.}
    \label{fig:axial}
\end{figure}

\subsubsection{Rotlet oriented in the transverse direction}
\label{subsubsec:TRANSVERSE}

\begin{figure}
    \centering
    \includegraphics[width=0.7\linewidth]{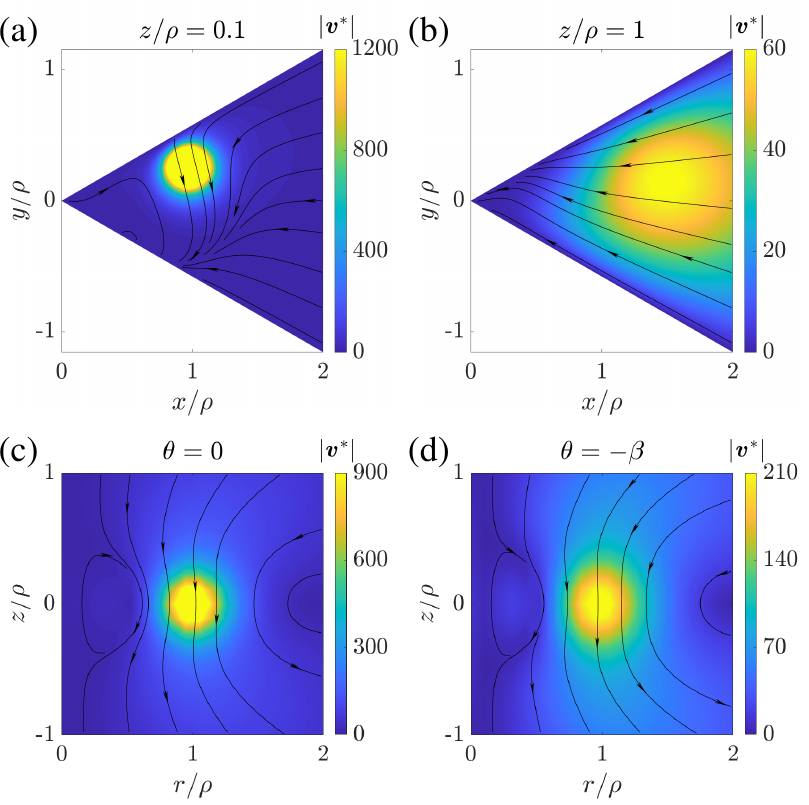}
    \caption{Streamlines and quiver plots of the viscous flow field in analogy to Fig.~\ref{fig:axial}, yet now induced by a radially oriented rotlet singularity pointing away from the tip of the wedge-like confinement. All other parameters are identical to those in Fig.~\ref{fig:axial},
    while these plots are generated from the results in Sec.~\ref{subsubsec:TRANSVERSE} with $\delta=0$.}
    \label{fig:radial}
\end{figure}

\begin{figure}
    \centering
    \includegraphics[width=0.7\linewidth]{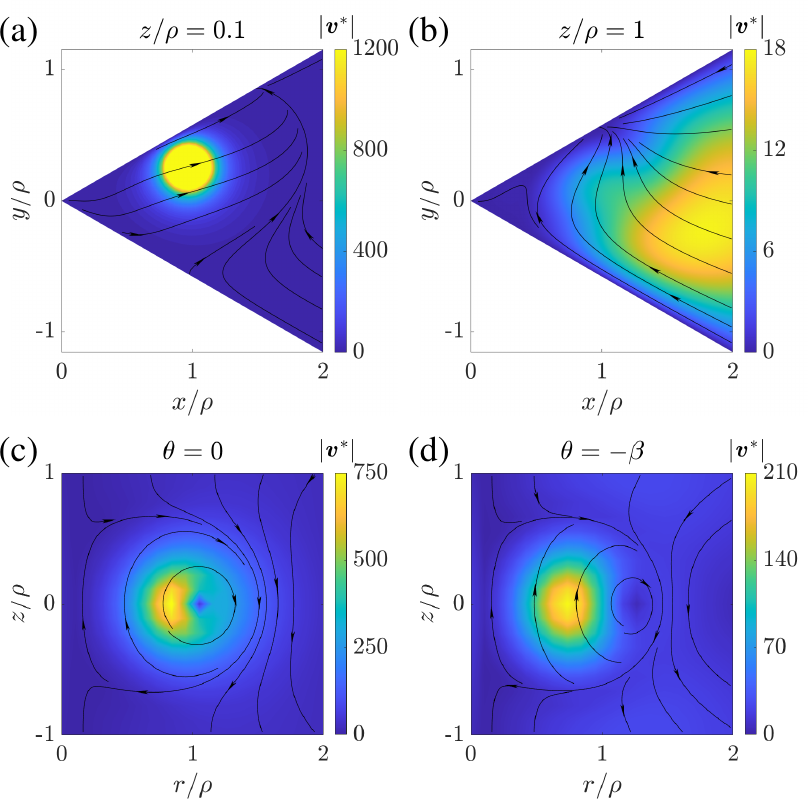}
    \caption{Similarly to Fig.~\ref{fig:axial}, streamlines and quiver plots of the viscous flow field are evaluated, yet now induced by an azimuthally oriented rotlet singularity, pointing in a direction perpendicular to the edge of the wedge and to the radial direction. All other parameters are identical to those in Fig.~\ref{fig:axial},
    whereas the present plots are generated from the results in Sec.~\ref{subsubsec:TRANSVERSE} with $\delta=\pi/2$.
    }
    \label{fig:azimuthal}
\end{figure}

We define $\delta$ as the angle between the torque vector and the radial direction, such that $\tan\delta = L_r / L_\theta$. Accordingly, $\delta = 0$ corresponds to purely radial torque, while $\delta = \pi/2$ corresponds to purely azimuthal torque.

The free-space components of the solution for a radial rotlet are given by
\begin{equation}
    \phi_r^\infty = \frac{z}{s^3}\, \sin(\theta-\beta-\delta) \, , \quad
    \phi_\theta^\infty = \frac{z}{s^3}\, \cos(\theta-\beta-\delta) \, , \quad
    \phi_z^\infty = -\frac{1}{s^3}
    \left( \rho\sin\theta+r\sin(\theta-\beta-\delta) \right) , \quad
    \phi_w^\infty = \frac{z}{s^3}\, \sin\delta \, .
\end{equation}
Corresponding expressions in FKL space are obtained as
\begin{align}
    \widetilde{\phi}_r^\infty &= -\sin(\theta-\beta-\delta) \,  
    \mathscr{T}_{i\nu} \left\{ \frac{\partial}{\partial z} \frac{1}{s} \right\} , 
    \qquad
    \widetilde{\phi}_\theta^\infty = -\cos(\theta-\beta-\delta) \, 
    \mathscr{T}_{i\nu} \left\{ \frac{\partial}{\partial z} \frac{1}{s} \right\} , 
    \notag \\
    \widetilde{\phi}_z^\infty &= \sin\delta \, \frac{\partial}{\partial\rho}  \mathscr{T}_{i\nu} \left\{ \frac{1}{s} \right\} -\cos\delta \sgn(\theta-\beta) \, \frac{1}{\rho} \,  \mathscr{T}_{i\nu} \left\{ \frac{\partial}{\partial a} \frac{1}{s} \right\}  , \quad
    \widetilde{\phi}_w^\infty = -\sin\delta \, \mathscr{T}_{i\nu} \left\{ \frac{\partial}{\partial z} \frac{1}{s} \right\}.
\end{align}
In this case, since $r\phi_r^\infty + z\phi_z^\infty + \rho \phi_w^\infty = 0$, Eq.~\eqref{eq:BCs_real} can be written in FKL space as
\begin{equation}
    \widetilde{\phi}_r^\infty + \widetilde{\phi}_r = 0 \, , \quad
    \widetilde{\phi}_z^\infty+\widetilde{\phi}_z = 0 \, , \quad
    \widetilde{\phi}_w^\infty+\widetilde{\phi}_w = 0 \, , \quad
    \frac{\partial  \widetilde{\phi}_r}{\partial\theta}
    + \frac{\partial}{\partial\theta}
    \, \mathscr{T}_{i\nu}
    \left\{ \frac{z}{r}\, \phi_z + \frac{\rho}{r}\, \phi_w \right\} - 2\left( \widetilde{\phi}_\theta^\infty + \widetilde{\phi}_\theta \right)= 0 \, , \qquad \text{at}~ \theta = \pm\alpha \, .
\end{equation}
We obtain
\begin{equation}
    \Delta_w = -\sin\delta \sh(\beta\nu) \sh\left( (\pi-\alpha)\nu\right) \csh(\alpha\nu) \, , \qquad
    \Delta_w^\dagger = -\sin\delta \ch(\beta\nu) \ch\left( (\pi-\alpha)\nu\right)\sch(\alpha\nu) \, ,
\end{equation}
and 
\begin{equation}
    \Lambda_z = \nu \cos\delta \ch(\beta\nu) \sh \left( (\pi-\alpha)\nu \right) \csh(\alpha\nu) \, , \qquad
    \Lambda_z^\dagger = \nu \cos\delta \sh(\beta\nu) \ch \left( (\pi-\alpha)\nu \right) \sch(\alpha\nu) \, .
\end{equation}
In addition, $\mathrm{H}_z = \Delta_w$ and $\mathrm{H}_z^\dagger = \Delta_w^\dagger$.

We express the solution for $j\in \{x,y\}$ in the form
\begin{equation}
    \Delta_j = \Gamma_\sigma \left( c_j \nu \sin(2\alpha) + d_j  \right) \, , \qquad\quad
    \Delta_j^\dagger = \Gamma_{-\sigma} \left( c_j^\dagger \nu \sin(2\alpha) + d_j^\dagger  \right) \, ,
\end{equation}
where, again, $\sigma = -$ for $j = x$ and $\sigma = +$ for $j = y$, and the expression of~$\Gamma_\pm^{-1}$ is given by Eq.~\eqref{eq:GAMMA}.
Defining $C=\cos(\beta+\delta)$ and $S=\sin(\beta+\delta)$, we obtain 
\begin{subequations}
\begin{align}
c_x &=  {S} \bigl( \lambda_{1}\ch(\pi \nu) - \sh(2 \alpha \nu)\sh(\pi \nu) \bigr) \sh(\beta\nu )
+  {C} \sin(2 \alpha)  \ch( \beta\nu) \sh(\pi \nu) \, , \\
c_y &= {C} \bigl( \sh(2 \alpha \nu) \sh(\pi \nu) - \lambda_{2} \ch(\pi \nu)  \bigr) \sh(\beta \nu)  
- {S}\sin(2\alpha)  \ch(\beta \nu) \sh(\pi \nu) \, ,\\
c_x^\dagger &= {S} \bigl( \sh(2 \alpha \nu)\sh(\pi \nu) - \lambda_{2} \ch(\pi \nu) \bigr) \ch(\beta \nu) 
+ {C} \sin(2\alpha) \sh(\beta \nu) \sh(\pi \nu) \, ,\\
c_y^\dagger &= {C} \left( \lambda_{1} \ch(\pi \nu)  - \sh(2 \alpha \nu) \sh(\pi \nu) \right) \ch(\beta \nu) 
-  {S} \sin(2 \alpha)  \sh(\beta \nu) \sh(\pi \nu) \, ,
\end{align}
\end{subequations}
and
\begin{subequations}
	\begin{align}
		d_x &= 
		\lambda_{1} {S} \ch(\pi \nu) \sh(2 \alpha \nu) \sh(\beta \nu) - 2 \ch(\alpha \nu) \sh(\pi \nu) \left( \delta_{1}^{-} {S} \ch(\alpha \nu) \sh(\beta \nu)  + {C} \sin(2 \alpha) \sh(\alpha \nu) \ch(\beta \nu)   \right), \\
		d_y &= \lambda_{2} {C} \ch(\pi \nu) \sh(2 \alpha \nu)  \sh(\beta \nu) - 2 \ch(\alpha \nu) \sh(\pi \nu) \left( \delta_{1}^{+} {C} \sh(\beta \nu) \ch(\alpha \nu) + {S} \sin(2 \alpha)  \sh(\alpha \nu) \ch(\beta \nu)  \right) , \\
		d_x^\dagger &= \lambda_{2} {S} \ch(\pi \nu)  \sh(2 \alpha \nu) \ch(\beta \nu) +2 \sh(\alpha \nu) \sh(\pi \nu) \left( {C} \sin(2 \alpha) \ch(\alpha \nu) \sh(\beta \nu)   -\delta_{2}^{+} {S} \sh(\alpha \nu) \ch(\beta \nu)    \right)  , \\
		d_y^\dagger &= \lambda_{1} {C} \ch(\pi \nu) \sh(2 \alpha \nu)  \ch(\beta \nu) +2 \sh(\alpha \nu) \sh(\pi \nu) \left( {S} \sin(2 \alpha)  \ch(\alpha \nu) \sh(\beta \nu)  - \delta_{2}^{-} {C} \sh(\alpha \nu) \ch(\beta \nu)  \right)  .
	\end{align}
\end{subequations}

\subsection{Solution in real space}

Having derived the solution of the flow problem in FKL space, we now apply the inverse transform in Eq.~\eqref{eq:inverse_FKL_transform} to recover the solution in real space. The integration with respect to the axial wavenumber $k$ can be carried out using tabulated integrals, leaving a single remaining integral over the radial wavenumber $\nu$, which is then evaluated numerically using standard routines.
Depending on the specific forms of the coefficients $A_j$ and $A_j^\dagger$ that define the solution for $\phi_j$, the resulting real-space solution will take correspondingly distinct forms.

For an axial rotlet, as well as for $j = z$ in the cases of radial and azimuthal rotlets, the coefficients in FKL space are given by Eq.~\eqref{eq:CoeffLamOmega}. In this setting, the solution can be written in the form
\begin{equation}
    \phi_j (r,\theta,z) = \int_0^\infty 
    \left( \frac{1}{\rho} \, \psi_j(\theta, \nu)
    + \xi_j(\theta, \nu) \, \frac{\partial}{\partial \rho} \right)
    \mathcal{K}_{i\nu}(r,z) \, \mathrm{d}\nu \, , 
    \label{eq:phi_parallel}
\end{equation}
with the kernel defined as
\begin{equation}
    \mathcal{K}_{i\nu}(r,z)
    = \left( \frac{2}{\pi} \right)^2 \int_0^\infty \cos(kz) K_{i\nu}(kr) K_{i\nu}(k\rho) \, \mathrm{d}k \, ,
    \label{eq:Kinu}
\end{equation}
and
\begin{equation}
    \psi_j(\theta, \nu) = q_l
    \left( \Lambda_j \sh(\theta\nu) + \Lambda_j^\dagger \ch(\theta\nu) \right) , \qquad
    \xi_j(\theta, \nu) = q_l
    \left( \mathrm{H}_j \sh(\theta\nu) + \mathrm{H}_j^\dagger \ch(\theta\nu) \right) .
    \label{eq:psi_and_xi}
\end{equation}
Here, $l \in \{\parallel, \perp\}$ specifies the direction of the point torque.

For transverse torque, the coefficients for $j \in \{x, y, w\}$ are given by Eq.~\eqref{eq:CoeffDelta}. In this setting, the solution takes the form
\begin{equation}
    \phi_j(r,\theta,z) = -\int_0^\infty \varphi_j(\theta,\nu) \, \frac{\partial }{\partial z} \, \mathcal{K}_{i\nu}(r,z) \, \mathrm{d}\nu \, , 
    \label{eq:phi_perp}
\end{equation}
where
\begin{equation}
    \varphi_j(\theta, \nu) = q_\perp \left( \Delta_j \sh(\theta\nu) + \Delta_j^\dagger \ch(\theta\nu) \right) .
    \label{eq:varphi}
\end{equation}
The improper integral in Eq.~\eqref{eq:Kinu} is convergent, and its value is given in classical textbooks as
\begin{equation}
    \mathcal{K}_{i\nu} (r,z) = 
    \left( \rho r\right)^{-\frac{1}{2}}
    P_{i\nu-\frac{1}{2}} (c) \, \sch(\pi\nu) \, ,
    \label{eq:Kinu_result}
\end{equation}
see Gradshteyn and Ryzhik~\cite[p.~719]{gradshteyn2014table}, Eq.~6.672, or Prudnikov \textit{et al.}~\cite[p.~390]{prudnikov1992integrals}, Eq.~2.16.36(2). Here, $P_n$ is the Legendre function of the first kind of degree $n$, with argument
\begin{equation}
    c = \frac{\rho^2+r^2+z^2}{2\rho r} \, .
\end{equation}

Depending on the direction of the applied point torque, the solution for the fluid velocity can be expressed as
\begin{equation}
    v_j(\bm{r}) = v_j^\infty(\bm{r})
    + \int_0^\infty \Psi_j(\bm{r}, \nu) \, \mathcal{K}_{i\nu}(r,z) \, \mathrm{d}\nu \, , \qquad j \in \{r,\theta,z\} \, ,
    \label{eq:final-solution}
\end{equation}
where the free-space velocity $v_j^\infty$ is defined in Eq.~\eqref{eq:unbounded}.
For axial point torque, we have $\phi_z = 0$ and $\xi_w = 0$.
The expressions for the operator $\boldsymbol{\Psi}$ are obtained by substituting the integral forms given in Eq.~\eqref{eq:phi_parallel} into the Papkovich–Neuber representation in Eq.~\eqref{eq:velocity_field}, yielding
\begin{equation}
    \Psi_r = \frac{\partial\Pi_\parallel}{\partial r} - 2
    \left( \frac{\psi_r}{\rho} + \xi_r \, \frac{\partial}{\partial\rho} \right) , \qquad
    \Psi_\theta = \frac{1}{r}\frac{\partial\Pi_\parallel}{\partial \theta} - 2
    \left( \frac{\psi_\theta}{\rho} + \xi_\theta \, \frac{\partial}{\partial\rho} \right) , \qquad
    \Psi_z = \frac{\partial\Pi_\parallel}{\partial z} \,  ,
\end{equation}
for axial point torque, and
\begin{equation}
    \Psi_r = \frac{\partial\Pi_\perp}{\partial r} + 2\varphi_r \, \frac{\partial}{\partial z} \, , \qquad
    \Psi_\theta = \frac{1}{r}\frac{\partial\Pi_\perp}{\partial \theta} + 2\varphi_\theta \, \frac{\partial}{\partial z} \, , \qquad
    \Psi_z = \frac{\partial\Pi_\perp}{\partial z}
    - 2 \left( \frac{\psi_z}{\rho} + \xi_z\, \frac{\partial}{\partial\rho} \right), 
\end{equation}
for transverse point torque.
The expressions for $\Pi_\parallel$ and $\Pi_\perp$ are obtained from Eq.~\eqref{eq:Pi} as
\begin{equation}
    \Pi_\parallel = r \left( \frac{\psi_r}{\rho} + \xi_r\, \frac{\partial}{\partial \rho} \right) + \psi_w \, , \qquad
    \Pi_\perp = z \left( \frac{\psi_z}{\rho} + \xi_z\, \frac{\partial}{\partial \rho} \right)
    - \left( r\varphi_r + \rho \varphi_w \right) \frac{\partial}{\partial z} \, .
\end{equation} 
Note that $\psi$, $\xi$, and $\varphi$ with subscripts $r$ and $\theta$ denote the projections of the corresponding quantities onto the radial and azimuthal directions in polar coordinates.

The infinite integrals in Eq.~\eqref{eq:final-solution} are evaluated numerically using the built-in command \texttt{NIntegrate} in the computer algebra system Mathematica~\cite{Mathematica}. Since the integrands involve several complicated functions, a sufficiently large working precision is required to ensure accurate results. 
In practice, a working precision of 100 is found to provide a reasonable balance between accuracy and computational efficiency.
As the integrands decay rapidly with increasing $\nu$, the integrals are truncated at a finite upper limit, which we take to be 100. Larger values slow down the numerical integration without notably affecting the results, as far as we observed.
Mathematica scripts used to generate these flow fields are freely available on Zenodo~\cite{zenodo2026RotletWedge}.

Based on these expressions, from Eq.~\eqref{eq:final-solution}, we can thus calculate the resulting flow field within the wedge-shaped confinement for any orientation of the applied rotlet. 
In Fig.~\ref{fig:axial}, we show a representative flow field generated by a torque singularity oriented in the axial direction, parallel to the edge of the wedge. In this example, the wedge has a semi–opening angle of $\alpha = \pi/6$, and the rotlet is positioned at $\beta = \alpha/2$.
Near the singularity, the flow resembles that of an unbounded fluid. At greater distances, however, this unbounded solution is distorted by the wedge-shaped confinement. The location of the velocity maximum in planes that do not contain the torque singularity shifts away from the wedge surface as the height above the rotlet increases.
In addition to the primary swirl induced by the rotlet, secondary swirling structures are present, though their velocities are significantly smaller.

For the same geometry as in Fig.~\ref{fig:axial}, Fig.~\ref{fig:radial} shows the flow field induced by a rotlet oriented radially away from the wedge edge. Here, we again observe that the location of the maximum velocity shifts away from the wedge surface with increasing height above the torque singularity. 

An example of the flow field generated by a torque applied in the azimuthal direction, that is, perpendicular both to the wedge edge and to the radial direction, is shown in Fig.~\ref{fig:azimuthal}. All other parameters are the same as those used for Fig.~\ref{fig:axial}.
As before, the velocity maximum is repelled by the boundaries and shifts toward the center of the wedge with increasing height above the torque. In the axial–radial plane, the flow is strongly distorted due to the wedge confinement. 

Overall, the wedge-shaped geometry has a pronounced effect on the resulting fluid flows compared to the unbounded geometry. In all cases, we find locally induced vortices, see, for example, Figs.~\ref{fig:axial}(c), \ref{fig:radial}(c), and \ref{fig:azimuthal}(c). Partly as a consequence, and together with the incompressibility of the fluid, we find that the flows significantly above the applied rotlet, see Figs.~\ref{fig:axial}(b), \ref{fig:radial}(b), and \ref{fig:azimuthal}(b), can show markedly different orientations than closer to the rotlet, see Figs.~\ref{fig:axial}(a), \ref{fig:radial}(a), and \ref{fig:azimuthal}(a), respectively.

\section{Solution in the planar-wall limit}
\label{sec:planar}

The problem of a point torque singularity acting near a planar no-slip wall can be recovered in the limit of a wedge semi–opening angle $\alpha = \pi/2$. Results for this geometry have previously been reported, for example, by Blake and Chwang~\cite{blake1974fundamental}; see also Spagnolie and Lauga~\cite{spagnolie12}. 
In that case, the coefficients take particularly simple forms. They are summarized in Tab.~\ref{tab:coeffs_pannar} for both axial and transverse torques.

\begin{table}[]
    \centering
   {
\renewcommand{\arraystretch}{1.1}
\setlength{\tabcolsep}{15pt}
\begin{tabular}{|c|c|c|}
    \hline
    Coefficients & Axial & Transverse \\
    \hline
    \hline
    $\Lambda_x$ & $-3\nu \cos\beta\ch(\beta\nu)$ & --- \\ 
    $\Lambda_x^\dagger$ & $-3\nu\cos\beta\sh(\beta\nu)$ & --- \\ 
    $\Lambda_y$ & $-\nu\sin\beta\ch(\beta\nu)$ & --- \\ 
    $\Lambda_y^\dagger$ & $-\nu\sin\beta\sh(\beta\nu)$ & --- \\ 
    $\Lambda_z$ & --- & $\nu\ch(\beta\nu) \cos\delta$ \\
    $\Lambda_z^\dagger$ & --- & $\nu\sh(\beta\nu) \cos\delta$ \\
    $\Lambda_w$ & $\nu\ch(\beta\nu)$ & --- \\ 
    $\Lambda_w^\dagger$ & $\nu\sh(\beta\nu)$ & --- \\ 
    \hline
    $\mathrm{H}_x$ & $-3\sin\beta\sh(\beta\nu)$ & --- \\
    $\mathrm{H}_x^\dagger$ & $-3\sin\beta\ch(\beta\nu)$ & --- \\
    $\mathrm{H}_y$ & $\cos\beta\sh(\beta\nu)$ & ---\\
    $\mathrm{H}_y^\dagger$ & $\cos\beta\ch(\beta\nu)$ & --- \\
    $\mathrm{H}_z$ & --- & $-\sh(\beta\nu) \sin\delta$ \\
    $\mathrm{H}_z^\dagger$ & --- & $-\ch(\beta\nu) \sin\delta$ \\
    \hline
    $\Delta_x$ & --- & $3\sh(\beta\nu) \sin(\beta+\delta)$ \\
    $\Delta_x^\dagger$ & --- & $3\ch(\beta\nu) \sin(\beta+\delta)$ \\
    $\Delta_y$ & --- & $-\sh(\beta\nu) \cos(\beta+\delta)$ \\
    $\Delta_y^\dagger$ & --- & $-\ch(\beta\nu) \cos(\beta+\delta)$ \\
    $\Delta_w$ & --- & $-\sh(\beta\nu) \sin\delta$ \\
    $\Delta_w^\dagger$ & --- & $-\ch(\beta\nu) \sin\delta$ \\
    \hline
    \hline
\end{tabular}
}
    \caption{Expressions for the coefficients defined in Eqs.~\eqref{eq:CoeffLamOmega} and \eqref{eq:CoeffDelta} in the planar wall limit, corresponding to $\alpha = \pi/2$.}
    \label{tab:coeffs_pannar}
\end{table}

\subsection{Axial torque}

Substituting the corresponding coefficients from Tab.~\ref{tab:coeffs_pannar} into Eq.~\eqref{eq:psi_and_xi}, we directly obtain
\begin{equation}
    \psi_x = -3q_\parallel \nu\cos\beta \sh \left( (\theta+\beta)\nu \right) \, , \quad
    \psi_y = -q_\parallel \nu\sin\beta \sh \left( (\theta+\beta)\nu \right) \, , \quad
    \psi_z = 0\, , \quad
    \psi_w = q_\parallel \nu \sh \left( (\theta+\beta)\nu\right) \, ,
\end{equation}
and 
\begin{equation}
    \xi_x = -3q_\parallel \sin\beta \ch \left( (\theta+\beta)\nu\right) \, , \quad
    \xi_y = q_\parallel \cos\beta \ch \left( (\theta+\beta)\nu\right) \, , \quad
    \xi_z = \xi_w = 0 \, .
\end{equation}
Using the integral presented in Appendix~C of Ref.~\onlinecite{daddi2025proc},
\begin{equation}
    \mathscr{L} = 
    \left( \rho r\right)^{-\frac{1}{2}}
    \int_0^\infty \ch(b\nu) \sch(\pi\nu) \, P_{i\nu-\frac{1}{2}}(c) \, \mathrm{d}\nu 
    = \frac{1}{\sqrt{r^2 + \rho^2 + 2\rho r \cos b + z^2}}
    \, , 
\end{equation}
for $ b \in [-\pi,\pi]$, if follows from Eq.~\eqref{eq:phi_parallel} that 
\begin{equation}
    \phi_x = -3q_\parallel \left( \frac{\cos\beta}{\rho} \frac{\partial}{\partial b} -\sin\beta\, \frac{\partial}{\partial\rho}\right) \mathscr{L} , \qquad
    \phi_y = q_\parallel \left( \cos\beta \, \frac{\partial}{\partial\rho} - \frac{\sin\beta}{\rho} \frac{\partial}{\partial b} \right) \mathscr{L} \, , \qquad
    \phi_w = \frac{q_\parallel}{\rho} \frac{\partial \mathscr{L}}{\partial b} \, ,
\end{equation}
evaluated at $b=\theta+\beta$.
Simplification yields
\begin{equation}
    \phi_x = \frac{3q_\parallel}{\overline{s}^3}
    \left( \rho\sin\beta-r\sin\theta \right) , \qquad
    \phi_y = -\frac{q_\parallel}{\overline{s}^3}
    \left( \rho\cos\beta+r\cos\theta \right) , \qquad
    \phi_w = \frac{q_\parallel}{\overline{s}^3} \, 
    r \sin(\theta+\beta) \, ,
    \label{eq:harmonic_funcs_planar_axial}
\end{equation}
where $\overline{s} = \mathscr{L} \left( b=\theta+\beta\right) = \sqrt{r^2 + \rho^2 + 2\rho r \cos(\theta + \beta) + z^2}$, representing the value of $s$ with $\beta$ substituted by $\pi - \beta$, the image position relative to the wall.

Using the expressions for the harmonic functions given in Eqs.~\eqref{eq:harmonic_funcs_planar_axial}, the components of the velocity field are first obtained in cylindrical coordinates. Upon transforming to Cartesian coordinates, the resulting velocity field is found to be in full agreement with the results reported by Blake and Chwang~\cite{blake1974fundamental}.

\subsection{Transverse torque}

Similarly, substituting the corresponding coefficients from Tab.~\ref{tab:coeffs_pannar} into Eqs.~\eqref{eq:psi_and_xi} and~\eqref{eq:varphi}, we readily obtain
\begin{equation}
    \varphi_x = 3q_\perp \sin(\beta+\delta) \ch\left( (\theta+\beta)\nu\right) \, , \quad
    \varphi_y = -q_\perp \cos(\beta+\delta) \ch\left( (\theta+\beta)\nu\right) \, , \quad
    \varphi_w = -q_\perp \sin\delta \ch\left( (\theta+\beta)\nu\right) \, ,
\end{equation}
and 
\begin{equation}
    \psi_z = q_\perp \nu \cos\delta \sh \left( (\theta+\beta)\nu\right) \, , \qquad
    \xi_z = -q_\perp \sin\delta \ch \left( (\theta+\beta)\nu\right) \, .
\end{equation}
It follows from Eqs.~\eqref{eq:phi_parallel} and~\eqref{eq:phi_perp} that
\begin{equation}
    \phi_x = -3q_\perp \sin(\beta+\delta) \,\frac{\partial\mathscr{L}}{\partial z} \, , \quad
    \phi_y = q_\perp \cos(\beta+\delta) \,\frac{\partial\mathscr{L}}{\partial z} \, , \quad
    \phi_z = q_\perp \left( \frac{\cos\delta}{\rho} \frac{\partial}{\partial b} -\sin\delta\, \frac{\partial}{\partial\rho}\right) \mathscr{L} , \quad
    \phi_w = q_\perp \sin\delta \, \frac{\partial\mathscr{L}}{\partial z} \, ,
\end{equation}
evaluated at $b=\theta+\beta$.
The final expressions of the harmonic functions are obtained as
\begin{equation}
    \phi_x = \frac{3q_\perp}{\overline{s}^3} \, z \sin(\beta+\delta) \, , \quad
    \phi_y = -\frac{q_\perp}{\overline{s}^3} \, z \cos(\beta+\delta) \, , \quad
    \phi_z = \frac{q_\perp}{\overline{s}^3}
    \left( \rho\sin\delta + r\sin(\theta+\beta+\delta) \right) , \quad
    \phi_w = -\frac{q_\perp}{\overline{s}^3} \, z\sin\delta \, .
\end{equation}

These results yield expressions for the velocity field that are in full agreement with those reported by Blake and Chwang~\cite{blake1974fundamental} after conversion from cylindrical to Cartesian coordinates.

\section{Hydrodynamic mobilities}
\label{sec:mobilities}

The solution for the flow field derived above can be used to evaluate how the confining wedge influences the motion of a small spherical colloidal particle of radius~$a$ subjected to an external torque. Due to hydrodynamic interactions with the no-slip boundaries, the particle will in general be displaced, and its rotational velocity will be affected. The leading-order term in the hydrodynamic coupling mobility is obtained by evaluating the image solution for the induced flow field at the position of the particle. This procedure yields an expression for the mobilities in terms of the small parameter $\epsilon = a/d$, where $d = \rho \sin(\alpha - \beta)$ and $a$ is the particle radius. 
Changes in rotational mobility when compared to the situation in an unbounded fluid are obtained by evaluating half the vorticity of the flow resulting from the image solution at the position of the rotating particle.

The translational and rotational velocities resulting from the hydrodynamic interactions with the boundaries are related to the hydrodynamic torque $\bm{L}$ acting on the particle through
\begin{equation}
    \bm{V} = \left. \bm{v}(\bm{r})-\bm{v}^\infty(\bm{r}) \right|_{\bm{r}=\bm{r}_0} \, , \qquad
    \boldsymbol{\Omega} = \left. \mu_0^{r} \bm{L}
    + \tfrac{1}{2} \, \boldsymbol{\nabla} \times 
    \left( \bm{v}(\bm{r})-\bm{v}^\infty(\bm{r}) \right) \right|_{\bm{r}=\bm{r}_0} \, .
    \label{eq:def_mobi}
\end{equation}
Here, $\mu_0^{r} = 1/(8\pi\eta a^3)$ denotes the bulk hydrodynamic rotational mobility in an unbounded fluid.

Written in a more general form, the grand mobility tensor relates the translational and angular velocities to the forces and torques acting on the particle via
\begin{equation}
    \begin{pmatrix}
        \bm{V} \\
        \boldsymbol{\Omega}
    \end{pmatrix}
    = 
     \begin{pmatrix}
       \bm{M}^{tt} & \bm{M}^{tr} \\
       \bm{M}^{rt} & \bm{M}^{rr}
    \end{pmatrix}
    \cdot
    \begin{pmatrix}
        \bm{F} \\
        \bm{L}
    \end{pmatrix}.
\end{equation}
Vice versa, the grand resistance tensor is defined via
\begin{equation}
   \begin{pmatrix}
        \bm{F} \\
        \bm{L}
    \end{pmatrix}
    = -
     \begin{pmatrix}
       \bm{R}^{tt} & \bm{R}^{tr} \\
       \bm{R}^{rt} & \bm{R}^{rr}
    \end{pmatrix}
    \cdot
     \begin{pmatrix}
        \bm{V} \\
        \boldsymbol{\Omega}
    \end{pmatrix} .
\end{equation}
We express the elements of the mobility tensor in the point-particle approximation as
\begin{equation}
    \bm{M}^{tt} = \mu_0^{t} \left( \bm{I} - \epsilon \bm{f}^{tt} \right) , \quad
    \bm{M}^{tr} = -\mu_0^{c} \, \epsilon^2 \bm{f}^{tr} \, , \quad
    \bm{M}^{rr} = \mu_0^{r} \left( \bm{I} - \epsilon^3 \bm{f}^{rr} \right) , 
\end{equation}
where $\bm{I}$ denotes the identity matrix and $\bm{M}^{rt}$ is the transpose of $\bm{M}^{tr}$, such that $M_{ij}^{rt}=M_{ji}^{tr}$. Moreover, $\mu_0^{t}=1/(6\pi\eta a)$ denotes the hydrodynamic translational mobility in a bulk fluid. $\mu_0^{c}=1/(6\pi\eta a^2)$ is used to scale the mobility coupling coefficients.
It follows that, to leading order, the translational mobilities scale as $\epsilon$, the coupling mobilities as $\epsilon^2$, while the corrections to the rotational mobilities scale as $\epsilon^3$.

Since $\bm{M}$ and $\bm{R}$ are inverses of each other, the elements of $\bm{R}$ can be expressed, to leading order in $\epsilon$, as
\begin{equation}
    \bm{R}^{tt} = \gamma_0^{t} \left( \bm{I} + \epsilon \bm{f}^{tt} \right) , \quad
    \bm{R}^{tr} = \gamma_0^{c} \, \epsilon^2 \bm{f}^{tr} \, , \quad
    \bm{R}^{rr} = \gamma_0^{r} \left( \bm{I} + \epsilon^3 \bm{f}^{rr} \right) . 
\end{equation}
$\bm{R}^{rt}$ is the transpose of $\bm{R}^{tr}$.
Here, $\gamma_0^t = 1/\mu_0^t$, $\gamma_0^r = 1/\mu_0^r$, and $\gamma_0 = 8\pi\eta a^2$ are used to scale the resistance coefficients.
For rotational–translational coupling in the hydrodynamic mobilities the scaling is $1/(6\pi\eta a^2)$, whereas for the corresponding resistance it must be $8\pi\eta a^2$ to ensure that the grand mobility and resistance tensors are exact inverses of one another.

Accordingly, when a torque is exerted on the particle, the velocities read
\begin{equation}
    \bm{V} = -\mu_0^{c} \,
    \epsilon^2 \bm{f}^{tr} \cdot \bm{L} \, , \qquad
    \boldsymbol{\Omega} = \mu_0^{r} 
    \left( \bm{I} - \epsilon^3 \bm{f}^{rr} \right) \cdot
    \bm{L} \, ,
\end{equation}
where
\begin{equation}
    \bm{f}^{tr} =
    \renewcommand{\arraystretch}{0.5}
    \begin{pmatrix}
        0 & 0 & a_{rz} \\
        0 & 0 & a_{\theta z} \\
        a_{zr} & a_{z\theta} & 0
    \end{pmatrix}
    , \qquad 
    \bm{f}^{rr} = 
    \begin{pmatrix}
        b_{rr} & b_{r\theta} & 0 \\
        b_{\theta r} & b_{\theta\theta} & 0 \\
        0 & 0 & b_{zz} 
    \end{pmatrix}  
\end{equation}
are dimensionless matrices, corresponding to the rotational--translational coupling mobilities and purely rotational mobilities, respectively.
By the symmetry of the mobility tensor, it follows that $b_{r\theta} = b_{\theta r}$.
Accordingly, we are left with determining eight components that fully characterize the response to the applied torque.
We note that the matrix $\bm{C}$ defined in Eq.~(5.2) of Sano and Hasimoto~\cite{sano1978effect} coincides with $\bm{f}^{rt}$, the transpose of $\bm{f}^{tr}$. Accordingly, we have verified that $t_{ij}=a_{ji}$, where $t_{ij}$ denote the elements of $\bm{C}$ as defined in Eq.~(5.3) of Sano and Hasimoto~\cite{sano1978effect}.

The following results are derived using Eq.~\eqref{eq:def_mobi} together with the solution for the flow field as given by Eq.~\eqref{eq:final-solution}.

\subsection{Hydrodynamic coupling mobilities}

\begin{figure}
    \centering
    \includegraphics[width=\linewidth]{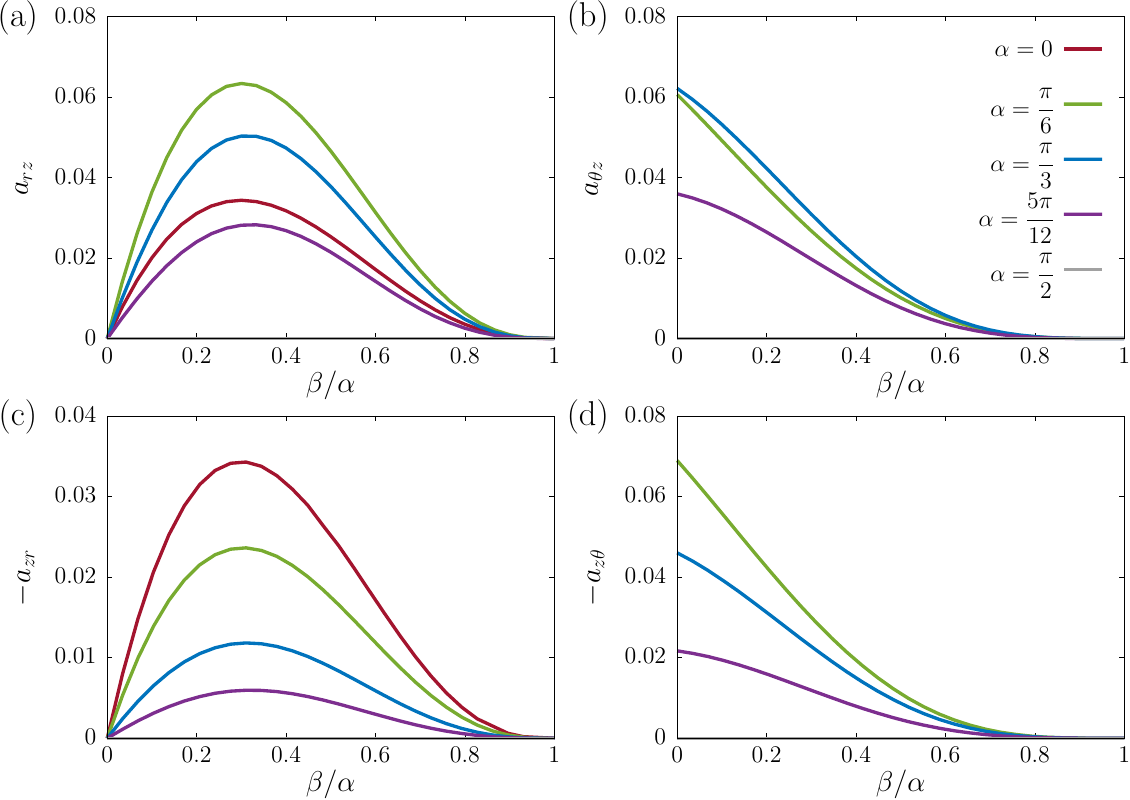}
    \caption{Variations of the leading-order corrections to the different components of the hydrodynamic coupling mobilities as functions of $\beta/\alpha$ for various values of the semi–opening angle $\alpha$ of the wedge. To leading order, the corrections vanish for $\alpha = \pi/2$ in panels (a) and (c), and for both $\alpha = 0$ and $\alpha = \pi/2$ in panels (b) and (d). Some curves are omitted to avoid overcrowding the figure. }
    \label{fig:coupling_mobi}
\end{figure}

The coefficients $a_{rz}$ and $a_{\theta z}$ associated with the axial torque can be written in the form
\begin{equation}
    a_{iz} = \frac{3}{64} \, \sin^2(\alpha-\beta) 
    \int_0^\infty
    \left( M_{i} \ch(\beta\nu) + N_{i} \sh(\beta\nu) \right) \sch(\pi\nu) \, \mathrm{d}\nu \, ,  
\end{equation}
for $i \in \{r,\theta\}$.
Here, we have defined
\begin{subequations}
    \begin{align}
    M_{r} &=
    \left( 12\Lambda_x^\dagger-\left(4\nu^2+7\right) \mathrm{H}_x^\dagger \right)\cos\beta
    +
    \left( 12\Lambda_y^\dagger-\left(4\nu^2+7\right) \mathrm{H}_y^\dagger \right)\sin\beta
    +4\Lambda_w^\dagger \, , \\
    N_{r} &=
    \left( 12\Lambda_x-\left(4\nu^2+7\right) \mathrm{H}_x \right)\cos\beta
    +
    \left( 12\Lambda_y-\left(4\nu^2+7\right) \mathrm{H}_y \right)\sin\beta
    +4\Lambda_w  \, , 
\end{align}
\end{subequations}
and 
\begin{subequations}
    \begin{align}
    M_{\theta} &= 4 \left( 
    \left( \left( \mathrm{H}_x-2\Lambda_x \right) \nu+2\Lambda_y^\dagger-\mathrm{H}_y^\dagger \right) \cos\beta
    + \,
     \left( \left( \mathrm{H}_y-2\Lambda_y \right) \nu-2\Lambda_x^\dagger+\mathrm{H}_x^\dagger \right) \sin\beta
     -2\nu \Lambda_w \right) \, , \\
    N_{\theta} &= 
    4\left( \left( \left( \mathrm{H}_x^\dagger-2\Lambda_x^\dagger \right) \nu+2\Lambda_y-\mathrm{H}_y \right) \cos\beta
    + 
     \left( \left( \mathrm{H}_y^\dagger-2\Lambda_y^\dagger \right) \nu-2\Lambda_x+\mathrm{H}_x \right) \sin\beta
     -2\nu \Lambda_w^\dagger \right) \, .
\end{align}
\end{subequations}

For the coefficients $a_{zr}$ and $a_{z\theta}$ associated with the transverse torque, the results are expressed in the form
\begin{equation}
        a_{zj} 
    = \frac{3}{64} \, \sin^2(\alpha-\beta)
    \int_0^\infty
    \left( 
    P_{zj}
    \ch(\beta\nu) 
    +
   Q_{zj}
    \sh(\beta\nu) \right) \sch(\pi\nu) \, \mathrm{d}\nu \, ,
\end{equation}
for $j\in\{r,\theta\}$, where
\begin{subequations}
    \begin{align}
    P_{zj} &=  4 \left( 2\Lambda_z^\dagger -\mathrm{H}_z^\dagger\right)  -\left( 4\nu^2+1 \right) \left( \Delta_x^\dagger \cos\beta + \Delta_y^\dagger \sin\beta + \Delta_w^\dagger \right) \Big|_{\delta = 0 \text{ for } j=r \text{ and } \delta= \frac{\pi}{2}  \text{ for } j=\theta\, 
    }
    , \\
    Q_{zj} &= 4 \left( 2\Lambda_z-\mathrm{H}_z \right) \, -\left( 4\nu^2+1 \right) \left( \Delta_x \cos\beta + \Delta_y \sin\beta + \Delta_w \right) \Big|_{\delta = 0 \text{ for } j=r \text{ and } \delta= \frac{\pi}{2}  \text{ for } j=\theta\, 
    }
    .  
\end{align}
\end{subequations}

In Fig.~\ref{fig:coupling_mobi}, we present the leading-order contributions to the various components of the hydrodynamic coupling mobilities as functions of $\beta/\alpha$ for different values of the wedge semi-opening angle $\alpha$. The components $a_{rz}$ and $-a_{zr}$ [panels (a) and (c)] exhibit a nonmonotonic dependence. They reach a maximum at an intermediate value $\beta/\alpha \approx 0.3$ and vanish at $\beta=0$ and $\beta=\alpha$. For a semi-opening angle of the wedge $\alpha=\pi/2$, corresponding to a planar wall, these components become zero. The limit $\alpha\to 0$ corresponds to the case of two infinitely extended parallel walls as addressed, for instance, by Swan and Brady~\cite{swan10}. In particular, for $\alpha \to 0$ and $\beta/\alpha = 1/2$, we find $a_{rz} = -a_{zr} \approx 0.025$, consistent with earlier results for a particle moving in a quarter-plane geometry. This value was first obtained by Faxén~\cite{faxen21} in his PhD dissertation and later improved by Wakiya~\cite{wakiya1956}, see also Happel and Brenner~\cite{happel12} [Eq.~(7--4.26), p.~326] and Ho and Leal~\cite{ho74}. In this case, the direction of translation is opposite to the one expected for a sphere in contact with the nearer wall
\footnote{This result follows from Eq.~(3.22b) by noting that $\kappa/\epsilon = 1/4$ and using the value $L_A = 0.270$ from Tab.~2. Substituting these values gives $(3/2)(1/4)^2 \times 0.270 \approx 0.025$. The factor $3/2 = 12/8$ arises from the different prefactors employed in their respective formulations.
}.

As previously noted, since the grand mobility and resistance tensors are inverses of each other, we have $a_{zr} = t_{rz}$, $a_{z\theta} = t_{\theta z}$, $a_{rz} = t_{zr}$, and $a_{\theta z} = t_{z\theta}$, where the rotational--translational coupling components $t_{ij}$ are defined by Sano and Hasimoto~\cite{sano1978effect} in their Eqs.~(3.12) and (4.8). Accordingly, panels (a), (b), (c), and~(d) of Fig.~\ref{fig:coupling_mobi} correspond to their Figs.~10, 11, 3, and~4, respectively~\cite{sano1978effect}.
Overall our results are in agreement with those reported by earlier studies.

Conversely, the components $a_{\theta z}$ and $-a_{z\theta}$ [panels (b) and (d) in Fig.~\ref{fig:coupling_mobi}] decrease monotonically with increasing $\beta/\alpha$, reaching their maximum values at $\beta=0$ and vanishing at $\beta=\alpha$. 
It is worth noting that, to leading order, no coupling occurs for $\alpha=0$ or $\alpha=\pi/2$. To quantify the translational motion induced by the rotating particle in these cases, one must include the Laplacian term in Faxén’s law in Eq.~\eqref{eq:def_mobi}, which introduces a correction that scales as $\epsilon^4$.

To better visualize the meaning of these mobility coefficients, we consider a simple set-up in which a particle is located on the bisector of the wedge and is subject to a torque. This corresponds to setting $\beta=0$, so that the radial and azimuthal directions coincide with the $x$ and $y$ directions, respectively. For a positive axial torque applied along the $z$ direction, the particle undergoes translational motion along the negative $y$ direction, indicated by $a_{rz}=0$ and $a_{\theta z}>0$. The magnitude of the translational velocity depends on the semi–opening angle $\alpha$, and there exists an intermediate angle at which this magnitude is maximal. When the particle is subject to a radial torque, it remains stationary, as expected, which is mathematically corroborated by the fact that $a_{zr}=0$ in this case. For an azimuthal torque applied in the positive direction, the particle undergoes translation along the positive $z$ direction, since $a_{z\theta}<0$. Likewise, there exists an optimal semi–opening angle for which the magnitude of the induced translational velocity is maximal.

It is worth noting that the observed behavior is counterintuitive in that the direction of motion induced by an externally applied torque is opposite to what one might expect. A similar effect has been reported for motion near a fluid–fluid interface~\cite{lee79}, where the leading correction, scaling as $\epsilon^2$, is directed opposite to intuitive expectations and opposite to that for a sphere near a solid flat wall, for which the leading-order contribution instead scales as $\epsilon^4$.

\subsection{Hydrodynamic rotational mobilities}

\begin{figure}
    \centering
    \includegraphics[width=\linewidth]{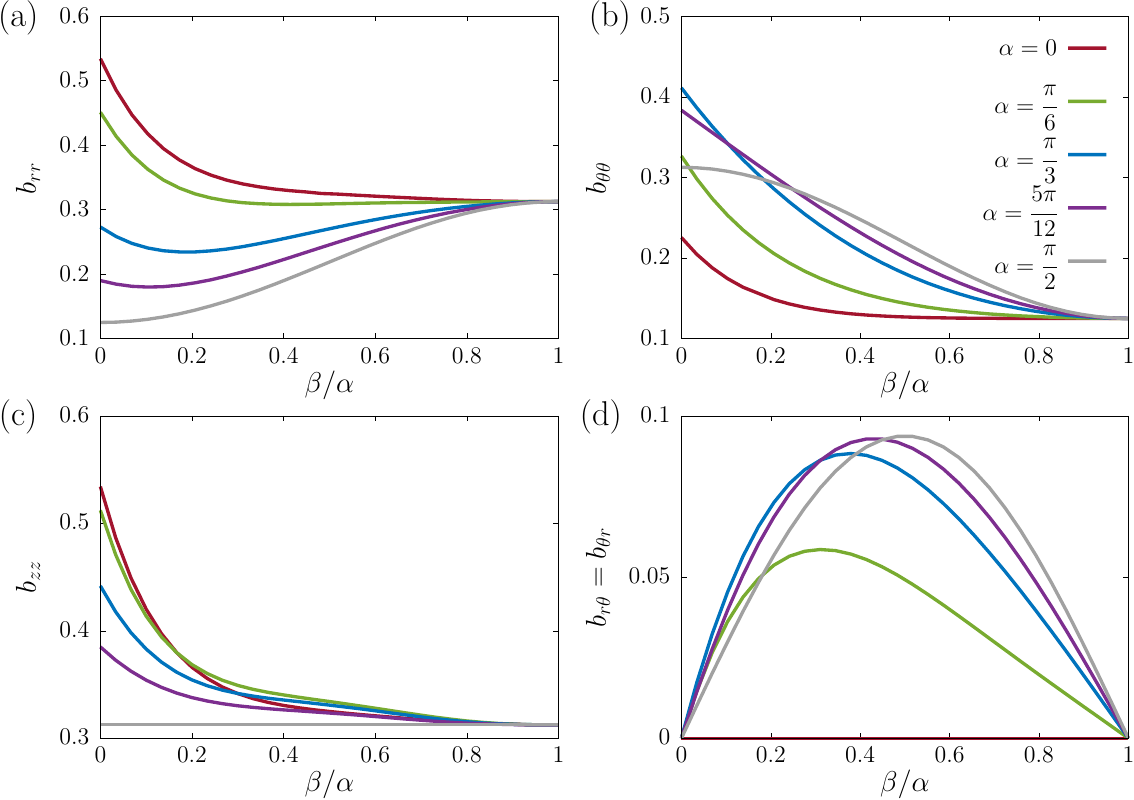}
    \caption{Variations of the leading-order corrections to the rotational mobilities as functions of $\beta/\alpha$ for various values of~$\alpha$. To leading order, the corrections vanish for the off-diagonal components $b_{r\theta} = b_{\theta r}$ [panel (d)] for $\alpha = 0$.
    Some curves are omitted for clarity. 
 }
    \label{fig:rot_mobi}
\end{figure}

Only one component is associated with axial torque, namely $b_{zz}$, and it is obtained as
\begin{equation}
    b_{zz} = \frac{1}{16}\, \sin^3(\alpha-\beta)
    \int_0^\infty \left( M_z \ch(\beta\nu) + N_z \sh(\beta\nu) \right) \sch(\pi\nu) \, \mathrm{d}\nu \, , 
\end{equation}
where
\begin{subequations}
    \begin{align}
    M_z &= \left( 4\nu\left( \mathrm{H}_x-2\Lambda_x \right)-4\Lambda_y^\dagger + \left( 4\nu^2+3\right) \mathrm{H}_y^\dagger \right)\cos\beta
    + \left( 4\nu \left( \mathrm{H}_y-2\Lambda_y \right) + 4\Lambda_x^\dagger - \left( 4\nu^2+3 \right) \mathrm{H}_x^\dagger \right) \sin\beta \, , \\
    N_z &= \left( 4\nu\left( \mathrm{H}_x^\dagger-2\Lambda_x^\dagger \right)-4\Lambda_y + \left( 4\nu^2+3\right) \mathrm{H}_y \right)\cos\beta
    + \left( 4\nu \left( \mathrm{H}_y^\dagger-2\Lambda_y^\dagger \right) + 4\Lambda_x - \left( 4\nu^2+3 \right) \mathrm{H}_x \right) \sin\beta \, ,
\end{align}
\end{subequations}
The four remaining components are associated with transverse torque and can be expressed as
\begin{equation}
    b_{ij} = \frac{1}{16} \, \sin^3(\alpha-\beta)
    \int_0^\infty \left( P_{ij} \ch(\beta\nu) + Q_{ij} \sh(\beta\nu) \right) \sch(\pi\nu) \, \mathrm{d}\nu \, , 
\end{equation}
where $i,j \in \{r,\theta\}$.
In addition,
\begin{subequations}
    \begin{align}
    P_{rj} &= \left( 4\nu^2+1\right) \left( \Delta_x^\dagger\sin\beta-\Delta_y^\dagger\cos\beta \right) + 4\nu \left( 2\Lambda_z-\mathrm{H}_z \right) \Big|_{\delta = 0 \text{ for } j=r \text{ and } \delta= \frac{\pi}{2}  \text{ for } j=\theta\, 
    }
    \, , \\
    Q_{rj} &= \left( 4\nu^2+1\right) \left( \Delta_x\sin\beta -\Delta_y\cos\beta\right) + 4\nu \left( 2\Lambda_z^\dagger-\mathrm{H}_z^\dagger \right) \Big|_{\delta = 0 \text{ for } j=r \text{ and } \delta= \frac{\pi}{2}  \text{ for } j=\theta\, 
    }
     \, , 
\end{align}
\end{subequations}
and 
\begin{subequations}
    \begin{align}
    P_{\theta j} &= \left( 4\nu^2+1\right) \left( \Delta_x^\dagger\cos\beta + \Delta_y^\dagger\sin\beta \right)+4\Lambda_z^\dagger- \left( 4\nu^2+3\right) \mathrm{H}_z^\dagger ~\Big|_{\delta = 0 \text{ for } j=r \text{ and } \delta= \frac{\pi}{2}  \text{ for } j=\theta\, 
    }
    \, , \\
    Q_{\theta j} &= \left( 4\nu^2+1\right) \left( \Delta_x\cos\beta + \Delta_y\sin\beta \right)\, +4\Lambda_z- \left( 4\nu^2+3\right) \mathrm{H}_z ~\Big|_{\delta = 0 \text{ for } j=r \text{ and } \delta= \frac{\pi}{2}  \text{ for } j=\theta\, 
    }
     \, .
\end{align}
\end{subequations}

In Fig.~\ref{fig:rot_mobi}, we show the variations of the leading-order corrections to the rotational mobilities as functions of the ratio $\beta/\alpha$ for different values of $\alpha$. For the cases $\alpha = \pi/2$, corresponding to a planar wall, the previously known rotational mobilities are recovered. The same applies to the limit $\beta \to \alpha$ at positions finitely distanced from the edge of the wedge, when the presence of only one of the two planar boundaries dominates. Then, for rotation about an axis \textit{perpendicular} to the wall, the correction is $1/8 = 0.125$. This case corresponds to $b_{rr}$ and $\beta \to 0$ when $\alpha = \pi/2$  [panel~(a)], or to $b_{\theta\theta}$ when $\beta \to \alpha$ for any value of $\alpha$ [panel~(b)]. For rotation about an axis \textit{parallel} to the wall, the correction is $5/16 = 0.3125$. It corresponds to $b_{\theta\theta}$ and $\beta \to 0$ when $\alpha = \pi/2$, to $b_{rr}$ when $\beta \to \alpha$ for any value of $\alpha$, or to~$b_{zz}$ for $\alpha = \pi/2$ across the full range of $\beta$ [panel~(c)].

In the limit $\alpha \to 0$ and $\beta \to 0$, we recover the well-known rotational mobilities for a particle in the midplane between two parallel walls. For rotation about an axis \textit{perpendicular} to the walls, the leading-order correction is~\cite{daddi2021steady}
\begin{equation}
b_\perp \,=\, \frac{1}{32} \left( \zeta\left(3, \frac{1}{2} \right) - \zeta(3) \right) \,\approx \, 0.2254 \,,
\end{equation}
where $\zeta(s)$ is the Riemann zeta function and $\zeta(s,a)$ is the Hurwitz zeta function. This value corresponds to $b_{\theta\theta}$ in the limits $\alpha \to 0$ and $\beta \to 0$ [panel~(b)].
For rotation about an axis \textit{parallel} to the walls, the leading-order correction is obtained as
\begin{equation}
b_\parallel \,=\, \frac{1}{8} \left( \zeta(3) + \int_0^\infty \frac{k^2 \, \mathrm{d}k}{k + \sh k} \right) \,\approx\, 0.5342 \, .
\end{equation}
This value is recovered for $b_{rr}$ and $b_{zz}$ in the same limits, $\alpha \to 0$ and $\beta \to 0$ [panels (a) and~(c)].

To gain further insight into the meaning of these mobility coefficients, we again consider, as a simple example, a particle located on the wedge bisector, so that $\beta=0$. Unlike the coupling mobility, which is absent in a bulk fluid and for which the wedge geometry induces translation through hydrodynamic coupling, the effect of the wedge on rotational motion merely tends to slow down the particle rotation. Moreover, since the correction scales as $\epsilon^3$, this effect is even weaker.
For a radial torque, the particle undergoes rotation only along the radial direction to this order, because $b_{\theta r}=0$ in this case and $b_{z r}=0$. Smaller opening angles of the wedge lead to larger corrections and therefore to slower rotations; see Fig.~\ref{fig:rot_mobi}(a). For an azimuthal torque, we again expect purely azimuthal rotation to this order, because $b_{r\theta}=0$ in this case and $b_{z\theta}=0$. Yet, the dependence on the opening angle of the wedge here is non-monotonic, with a maximal correction at an intermediate angle and a minimum reached in the limit $\alpha=0$; see Fig.~\ref{fig:rot_mobi}(b). Finally, for an axial torque, the dependence on the wedge angle is also non-monotonic, exhibiting a minimum in the planar-boundary limit corresponding to $\alpha=\pi/2$; see Fig.~\ref{fig:rot_mobi}(c). There is no coupling to the other directional components to the considered order. 
Overall, the dynamics is strongly dependent on the geometry. As we have indicated, accurate predictions of the mobilities require a systematic quantification.

\section{Conclusions}
\label{sec:concl}

Summarizing, we here derived analytical expressions for low-Reynolds-number fluid flows induced by a localized torque in a wedge-shaped geometry. No-slip surface conditions are applied at the surfaces of the wedge-shaped confinement. To perform the mathematical derivation, we rely on the Fourier-Kontorovich-Lebedev transform. From the illustration of resulting flow fields, we observe the expected distortion of the resulting fluid flows and vortices by the wedge-shaped geometry. 

Wedge-shaped geometries of low-Reynolds-number flows play a significant role, for instance, in microfluidic devices, as outlined above~\cite{yang2018wedge}. In that example, separation of mixtures of biological cells was identified as an application of wedge-shaped channels. On the opposite side, enhanced mixing between different components or fluids is a frequent topic in microfluidics \cite{chen2004topologic,lee2016passive,li2022review}. Introducing rotational components is one strategy to promote this effect \cite{janssen2009controlled,chong2018magnetic,owen2016rapid}. Our mathematical expressions provide a quick estimate for the resulting induced fluid flows when assuming the rotational probes as localized in space. We recently derived corresponding expressions for the geometry of two parallel, concentric disks \cite{daddi2021steady}. Here, we provide them for the case of a rigid wedge. 

The point-particle approximation employed in the present work is generally valid in the far-field limit, when the particle is sufficiently distant from the nearest interface. A systematic assessment of its accuracy would require either exact analytical solutions for fully extended particles or detailed numerical simulations. Nevertheless, based on our previous experience with planar and spherical interfaces, our approximation—despite its simplicity—often provides remarkably accurate predictions of hydrodynamic mobilities, even at distances comparable to the particle diameter, which highlights its usefulness for practical calculations, qualitative estimates at smaller distances, and theoretical analyses~\cite{daddi16, daddi17c, daddi2018creeping}.
We note that our theory may exhibit quantitative deviations before near-field lubrication forces become significant. Nevertheless, when the particle approaches one of the two flat surfaces of the wedge, at sufficient distance from the tip of the wedge and the other surface, lubrication interactions with this one surface dominate. In this case, the approximation valid for one flat surface provides a reasonable leading-order description.
In future work, it would be interesting to study the problem using exact analytical frameworks, such as bipolar coordinates or related methods.

One step further, we relied on the derived solution for the flow induced by the rotlet to calculate the lowest-order hydrodynamic mobilities in this geometry. They express how objects small in size when compared to the distance from the boundaries are displaced and rotated when they exert a torque on the surrounding fluid. Such motion results from the hydrodynamic interactions with the no-slip boundaries of the wedge-shaped confinement. As a consequence, transport can be induced by imposed rotations, in contrast to the case of an unbounded fluid in the absence of any boundaries.

\begin{acknowledgments}
J.M.\ and A.M.M.\ thank the Deutsche Forschungs\-ge\-mein\-schaft (German Research Foundation, DFG) for support through project no.\ 541972050 (DFG reference no.\ ME 3571/12-1). 
E.T.\ acknowledges support from the Engineering and Physical Sciences Research Council (EPSRC) under grant no.~EP/W027194/1.
\end{acknowledgments}

\section*{Data Availability Statement}
Mathematica scripts that compute the flow field at a specified coordinate are available on the Zenodo data repository at \url{https://zenodo.org/records/18350815}.
All other data are available from the corresponding author upon reasonable request.

\appendix

\section*{Proof of the FKL properties}

To derive Eqs.~\eqref{eq:FKL-property-diff-R} and \eqref{eq:FKL-property-division-by-r} we make use of the fact that these transforms are unaffected by the Fourier transform, together with a number of standard recurrence relations for the  modified Bessel functions. These are
\begin{equation}
    \frac{K_{i\nu}(|k|r)}{r} = 
    \frac{|k|}{2i\nu} \left(
    K_{i\nu+1}(|k|r) - K_{i\nu-1}(|k|r) \right) 
    \label{eq:bessel_recurrence_1}
\end{equation}
and
\begin{equation}
    \frac{\partial}{\partial r} K_{i\nu}(|k|r) = -|k| K_{i\nu+1} (|k|r) + 
    \frac{i\nu}{r} \, K_{i\nu} (|k|r) \, .
     \label{eq:bessel_recurrence_2}
\end{equation}
From the recurrence relation given by Eq.~\eqref{eq:bessel_recurrence_1}, we readily arrive at
\begin{equation}
     \mathscr{T}_{i\nu} \left\{ \frac{f}{r} \right\}
    = \frac{|k|}{2i\nu} \left( \mathscr{T}_{i\nu+1} \{f\} - \mathscr{T}_{i\nu-1} \{f\} \right) . 
    \label{eq:FKL_f_r_appendix}
\end{equation}
Integration by parts and the recurrence relation given by Eq.~\eqref{eq:bessel_recurrence_2} lead to
\begin{equation}
    \hspace{-0.2cm}
    \mathscr{T}_{i\nu} \left\{ \frac{\partial f}{\partial r} \right\} =
    |k| \left( \mathscr{T}_{i\nu+1} \{f\} 
    -\left( i\nu-1 \right)
     \mathscr{T}_{i\nu} \left\{ \frac{f}{r} \right\} 
    \right).
\end{equation}
Finally, using Eq.~\eqref{eq:bessel_recurrence_1}, we find
\begin{equation}
    \mathscr{T}_{i\nu} \left\{ \frac{\partial f}{\partial r} \right\}
    = \frac{|k|}{2i\nu} 
    \bigl( (i\nu+1) \mathscr{T}_{i\nu+1} \{f\}
    +  (i\nu-1) \mathscr{T}_{i\nu-1} \{f\}
    \bigr) .
\end{equation}

The FKL transform of $(z/r)f$ is more involved, because the Fourier transform is affected in this case.
Differentiating the FKL transform of $f/r$ from Eq.~\eqref{eq:FKL_f_r_appendix} with respect to $k$, we obtain
\begin{equation}
ik \mathscr{T}_{i\nu} \left\{ \frac{z}{r} \, f\right\}
=   \left( k\, \frac{\partial}{\partial k} 
    - i\nu \right) \mathscr{T}_{i\nu} \left\{ \frac{f}{r} \right\}
    +|k| \mathscr{T}_{i\nu+1} \{ f\} \, . \notag 
\end{equation}
Together with Eq.~\eqref{eq:FKL_f_r_appendix}, which yields the FKL transform of $f/r$, we arrive at
\begin{equation}
    \mathscr{T}_{i\nu} \left\{ \frac{z}{r} \, f \right\} 
    = {} -\frac{\sgn k}{2\nu} 
    \left(
    \left(
    i\nu + 1 + k\, \frac{\partial}{\partial k} \right) \mathscr{T}_{i\nu+ 1} \left\{ f\right\}
    + \left(
    i\nu - 1 - k\, \frac{\partial}{\partial k} \right) \mathscr{T}_{i\nu-1} \left\{ f\right\}
    \right) .
    \label{eq:FKL_rzf}
\end{equation}

%


\end{document}